\documentclass[aps,prd,twocolumn,floatfix,superscriptaddress,preprintnumbers,tightenlines,showpacs,showkeys,nofootinbib]{revtex4-1}
\usepackage[utf8]{inputenc}
\usepackage[colorlinks=true,citecolor=blue,linkcolor=blue]{hyperref}
\usepackage[normalem]{ulem}
\usepackage{amsmath,amssymb, mathrsfs}
\usepackage{epsfig}
\usepackage{graphicx}               
\usepackage{url}
\usepackage{color}
\usepackage{slashed}
\usepackage{multirow}
\usepackage{placeins}
\usepackage[dvipsnames]{xcolor}
\usepackage{epstopdf}
\usepackage{soul}
\usepackage{tikz}
\usepackage[capitalise, english]{cleveref}
\usepackage{siunitx}
\usepackage{xspace}
\usepackage{booktabs}
\usetikzlibrary{trees}
\usetikzlibrary{decorations.pathmorphing}
\usetikzlibrary{decorations.markings}

\newcommand\myshade{80}
\colorlet{mylinkcolor}{ForestGreen}
\colorlet{mycitecolor}{Red}
\colorlet{myurlcolor}{violet}

\hypersetup{
  linkcolor  = mylinkcolor!\myshade!black,
  citecolor  = mycitecolor!\myshade!black,
  urlcolor   = myurlcolor!\myshade!black,
  colorlinks = true
}

\definecolor{jblue}{RGB}{20,50,100}
\definecolor{npurple}{RGB} {153, 51, 204}
\definecolor{wred}{RGB}{217,0,56}
\definecolor{white}{RGB}{255,255,255}
\definecolor{korange}{RGB}{235, 80,  43}
\definecolor{korange2}{RGB}{245, 100,  63}
\definecolor{kyelloworange}{RGB}{255, 210,  110}
\definecolor{kyelloworange2}{RGB}{240, 170,  90}
\definecolor{kred}{RGB}{204,  102, 153}
\definecolor{kpurple}{RGB}{153,  61, 190}
\definecolor{kpurplelight}{RGB}{213,  161, 230}

\allowdisplaybreaks

\usepackage{tikz,xcolor,hyperref}

\definecolor{lime}{HTML}{A6CE39}
\DeclareRobustCommand{\orcidicon}{\hspace{-1mm}
	\begin{tikzpicture}
	\draw[lime, fill=lime] (0,0) 
	circle [radius=0.16] 
	node[white] {{\fontfamily{qag}\selectfont \tiny \,ID}};
	\draw[white, fill=white] (-0.0525,0.095) 
	circle [radius=0.007];
	\end{tikzpicture}
	\hspace{-3mm}
}

\foreach \x in {A, ..., Z}{\expandafter\xdef\csname orcid\x\endcsname{\noexpand\href{https://orcid.org/\csname orcidauthor\x\endcsname}
			{\noexpand\orcidicon}}
}

\begin{document}

\title{Enhanced Muonization by Active-Sterile Neutrino Mixing in Protoneutron Stars}

\author{Anupam Ray\orcidA{}}
\email{anupam.ray@berkeley.edu}
\affiliation{Department of Physics, University of California Berkeley, Berkeley, California 94720, USA}
\affiliation{School of Physics and Astronomy, University of Minnesota, Minneapolis, MN 55455, USA}

\author{Yong-Zhong Qian\orcidB{}}
\email{qianx007@umn.edu}
\affiliation{School of Physics and Astronomy, University of Minnesota, Minneapolis, MN 55455, USA}

\date{\today}
\preprint{N3AS-24-013}

\begin{abstract}
We study $\nu_\mu$-$\nu_s$ and $\bar\nu_\mu$-$\bar\nu_s$ mixing in the protoneutron star (PNS) 
created in a core-collapse supernova (CCSN). We point out the importance of the feedback on the general composition of the PNS
in addition to the obvious feedback on the $\nu_\mu$ lepton number. We show that for our adopted mixing parameters
$\delta m^2\sim 10^2$~keV$^2$ and $\sin^2 2\theta$ consistent with the current constraints,
sterile neutrino production is dominated by the Mikheyev–Smirnov–Wolfenstein conversion of $\bar\nu_\mu$ into $\bar\nu_s$ 
and that the subsequent escape of $\bar\nu_s$ increases the $\nu_\mu$ lepton number, which in turn enhances muonization
of the PNS primarily through $\nu_\mu+n\to p+\mu^-$. While these results are qualitatively robust, their quantitative effects on the dynamics and active neutrino
emission of CCSNe should be evaluated by including $\nu_\mu$-$\nu_s$ and $\bar\nu_\mu$-$\bar\nu_s$ mixing in the simulations.
\end{abstract}

\maketitle
\preprint{}

\section{Introduction}  
Sterile neutrinos ($\nu_s$) associated with a vacuum mass eigenstate of keV-scale mass are a viable candidate 
for the mysterious dark matter \cite{Kusenko:2009up,Drewes:2016upu,Abazajian:2017tcc,Boyarsky:2018tvu,Dasgupta:2021ies}.
Because of their tiny mixing with active neutrinos ($\nu_\alpha$), they can decay
radiatively, $\nu_s \to \nu_\alpha + \gamma$, thereby allowing experimental probes through
the decay X-ray line. In fact, major experimental constraints on active-sterile neutrino mixing primarily 
come from such X-ray searches~\cite{Abazajian:2001vt,Boyarsky:2007ge,Ng:2019gch,Roach:2019ctw,Foster:2021ngm,Roach:2022lgo}. 
There are also other significant constraints from phase-space considerations (the Tremaine-Gunn bound)~\cite{PhysRevLett.42.407,Boyarsky:2008ju,Gorbunov:2008ka,DiPaolo:2017geq}, measurement of Lyman-$\alpha$ forests~\cite{Boyarsky:2008xj,Irsic:2017ixq,Zelko:2022tgf}, and terrestrial nuclear decay searches~\cite{Smith:2016vku,KATRIN:2018oow,Benso:2019jog,Balantekin:2023jlg}. 
Among many theoretical explorations, a number of studies have investigated the effects of active-sterile neutrino mixing
in core-collapse supernovae (CCSNe)
\cite{Shi:1993ee,Nunokawa:1997ct,Hidaka:2006sg,Hidaka:2007se,1993APh.....1..165R,2011PhRvD..83i3014R,Warren:2014qza,Warren:2016slz,2019PhRvD..99d3012A,2019JCAP...12..019S,Suliga:2020vpz,PhysRevD.106.015017,PhysRevD.108.063025,Carenza:2023old,Chauhan:2023sci}.
In this paper, we focus on the mixing between $\nu_{\mu}$ ($\bar\nu_{\mu}$) and $\nu_s$ ($\bar\nu_s$) 
with a vacuum mass-squared difference of $\delta m^2\sim 10^2$~keV$^2$ in protoneutron stars (PNSs) created in CCSNe.
In particular, we treat this mixing in the presence of muons. In contrast to Ref.~\cite{PhysRevD.106.015017}, which focused on 
the energy loss due to $\nu_{x}$-$\nu_s$ and $\bar\nu_{x}$-$\bar\nu_s$ ($x=\mu$, $\tau$) mixing and stated that the presence 
of muons makes little difference, we show that inclusion of muons leads to more complicated feedback effects for
$\nu_{\mu}$-$\nu_s$ and $\bar\nu_{\mu}$-$\bar\nu_s$ mixing than those for $\nu_{\tau}$-$\nu_s$ and $\bar\nu_{\tau}$-$\bar\nu_s$ mixing.

Due to the high densities and temperatures in the PNS, muon production or muonization 
(e.g., $e^-+\bar\nu_e\to\mu^-+\bar\nu_\mu$, $\nu_\mu+n\to p+\mu^-$) is energetically allowed and dynamically important 
for the neutrino-driven explosion \cite{2017PhRvL.119x2702B}. The dynamic effects of muonization can be understood qualitatively as follows.
For the conditions in the PNS, electrons are highly relativistic, and their contribution to the pressure is
$\approx 1/3$ of their energy density. Without muons, charge neutrality requires an equal number of electrons and protons. 
For a given initial distribution of protons, as some electrons are converted into muons, a fraction of the electron energy
is converted into the rest mass energy of muons, thereby reducing the net pressure. Consequently, the PNS contracts faster and
the resulting increase in neutrino luminosity enhances heating of the material outside the PNS, which helps the neutrino-driven explosion 
\cite{2017PhRvL.119x2702B}. The conditions in the PNS also facilitate conversion of $\bar\nu_\mu$ into $\bar\nu_s$ through
the Mikheyev–Smirnov–Wolfenstein (MSW) effect \cite{MikheyeSmirnov1985,1978PhRvD..17.2369W}, which tends to
increase the $\nu_\mu$ lepton number, thereby enhancing muonization mainly through the reaction $\nu_\mu+n\to p+\mu^-$.
While the corresponding dynamic effects must be investigated by incorporating $\nu_{\mu}$-$\nu_s$ and $\bar\nu_{\mu}$-$\bar\nu_s$ mixing
in CCSN simulations, our goal in this paper is to illustrate how such mixing enhances muonization in the PNS and to estimate
the potential enhancement.

The rest of the paper is organized as follows. In Section~\ref{sec:setup} we discuss the production of sterile neutrinos through $\nu_{\mu}$-$\nu_s$ and 
$\bar\nu_{\mu}$-$\bar\nu_s$ mixing in a PNS. Much of the treatment is similar to that of $\nu_{\tau}$-$\nu_s$ and 
$\bar\nu_{\tau}$-$\bar\nu_s$ mixing, and we closely follow the discussion in Ref.~\cite{PhysRevD.108.063025}.
In Section~\ref{sec:feedback} we discuss the treatment of the feedback effects of $\nu_{\mu}$-$\nu_s$ and $\bar\nu_{\mu}$-$\bar\nu_s$ 
mixing in the presence of muons. In Section~\ref{sec:example} we present example calculations to illustrate the effects of
such mixing on muonization in the PNS. In Section~\ref{sec:discuss} we discuss our results and give conclusions.

\section{Production of sterile neutrinos in a PNS}
\label{sec:setup}
We assume spherical symmetry, for which the conditions in the PNS are functions of radius $r$ and time $t$ only.
We focus on the region where all active neutrinos are diffusing and assume that all particles of concern, 
$n$, $p$, $e^\pm$, $\mu^\pm$, $\nu_e$, $\bar\nu_e$, $\nu_\mu$, $\bar\nu_\mu$, $\nu_\tau$, and $\bar\nu_\tau$,  have Fermi-Dirac energy distributions 
characterized by the same temperature $T$ but species-specific chemical potentials. For example, the $\nu_\mu$ and $\bar\nu_\mu$
energy distributions (number densities per unit energy interval per unit solid angle) are
\begin{subequations}
\begin{align}
     \frac{d^2n_{\nu_\mu}}{dEd\Omega}&=\frac{1}{(2\pi)^3}\frac{E^2}{\exp[(E-\mu_{\nu_\mu})/T]+1},\\
     \frac{d^2n_{\bar\nu_\mu}}{dEd\Omega}&=\frac{1}{(2\pi)^3}\frac{E^2}{\exp[(E+\mu_{\nu_\mu})/T]+1},
\end{align}
\end{subequations}
where $\mu_{\nu_\mu}$ is the $\nu_\mu$ chemical potential and we have used $\mu_{\bar\nu_\mu}=-\mu_{\nu_\mu}$.
The corresponding $\nu_\mu$ lepton number fraction (net number per baryon) is
\begin{align}
	Y_{\nu_\mu}=\frac{n_{\nu_\mu}-n_{\bar\nu_\mu}}{n_b}=\frac{T^3\eta_{\nu_\mu}}{6n_b}\left(1+\frac{\eta_{\nu_\mu}^2}{\pi^2}\right),
 \label{eq:ynumu}
\end{align}
where $\eta_{\nu_\mu}=\mu_{\nu_\mu}/T$ and $n_b$ is the baryon number density.

The treatment of $\nu_{\mu}$-$\nu_{s}$ and $\bar\nu_{\mu}$-$\bar\nu_{s}$ mixing would mirror that of 
$\nu_{\tau}$-$\nu_{s}$ and $\bar\nu_{\tau}$-$\bar\nu_{s}$ mixing (e.g., 
\cite{1993APh.....1..165R,2011PhRvD..83i3014R,2019PhRvD..99d3012A,2019JCAP...12..019S,PhysRevD.106.015017,PhysRevD.108.063025})
were there no muons. With muons, the effective potential for $\nu_{\mu}$-$\nu_{s}$ mixing is
\begin{equation}\label{eq:mswpot}
	V_{\nu} = \sqrt{2} G_F n_b  \left(-\frac{Y_n}{2}  + Y_{\mu} + Y_{\nu_e}+2Y_{\nu_{\mu}}\right),
\end{equation}
where $G_F$ is the Fermi constant, and 
$Y_{\alpha}$ ($\alpha = n$, $\mu$, $\nu_e$, and $\nu_{\mu}$) is the net number fraction for species $\alpha$. 
In Eq.~(\ref{eq:mswpot}), we have applied the constraint of charge neutrality, $Y_e+Y_\mu=Y_p$,
so the contributions from forward scattering of $\nu_\mu$ on protons and electrons do not appear explicitly.
We have also assumed that there is no net $\nu_\tau$ lepton number in the PNS (i.e., $Y_{\nu_\tau}=0$ and $\mu_{\nu_\tau}=0$).
Note that $V_{\nu}$, $n_b$, and $Y_{\alpha}$ are all functions of $r$ and $t$. For convenience, here and below, 
we usually suppress such radial and temporal dependence. The potential for $\bar\nu_{\mu}$-$\bar\nu_s$ mixing is
$V_{\bar\nu}=-V_{\nu}$.

The potential $V_\nu$ ($V_{\bar\nu}$) modifies the mixing angle between $\nu_{\mu}$ ($\bar{\nu}_{\mu}$) and
$\nu_s$ ($\bar\nu_s$) in the PNS. For $\nu_{\mu}$ and $\bar{\nu}_{\mu}$ with energy $E$, the effective mixing angles are given by
\begin{subequations}
\begin{align}
	\sin^22\theta_\nu&=\frac{\Delta^2\sin^22\theta}{(\Delta\cos2\theta-V_{\nu})^2+\Delta^2\sin^22\theta},\label{eq:thetanu}\\
	\sin^22\theta_{\bar\nu}&=\frac{\Delta^2\sin^22\theta}{(\Delta\cos2\theta-V_{\bar{\nu}})^2+\Delta^2\sin^22\theta},\label{eq:thetanubar}\,
\end{align}
\end{subequations}
where $\theta \ll1$ is the vacuum mixing angle and $\Delta=\delta m^2/(2E)$.
For the conditions in the PNS, $V_{\nu} < 0$ and $V_{\bar\nu} > 0$. Therefore, $\nu_{\mu}$-$\nu_s$ mixing is suppressed whereas
$\bar{\nu}_{\mu}$-$\bar{\nu}_s$ mixing can be enhanced by the MSW resonance. This resonance occurs when $\Delta\cos2\theta=V_{\bar\nu}$,
and defines a resonance energy $E_R=\delta m^2\cos2\theta/(2V_{\bar\nu})$.

\begin{figure}[!t]
	\centering
	\includegraphics[width=0.35\textwidth]{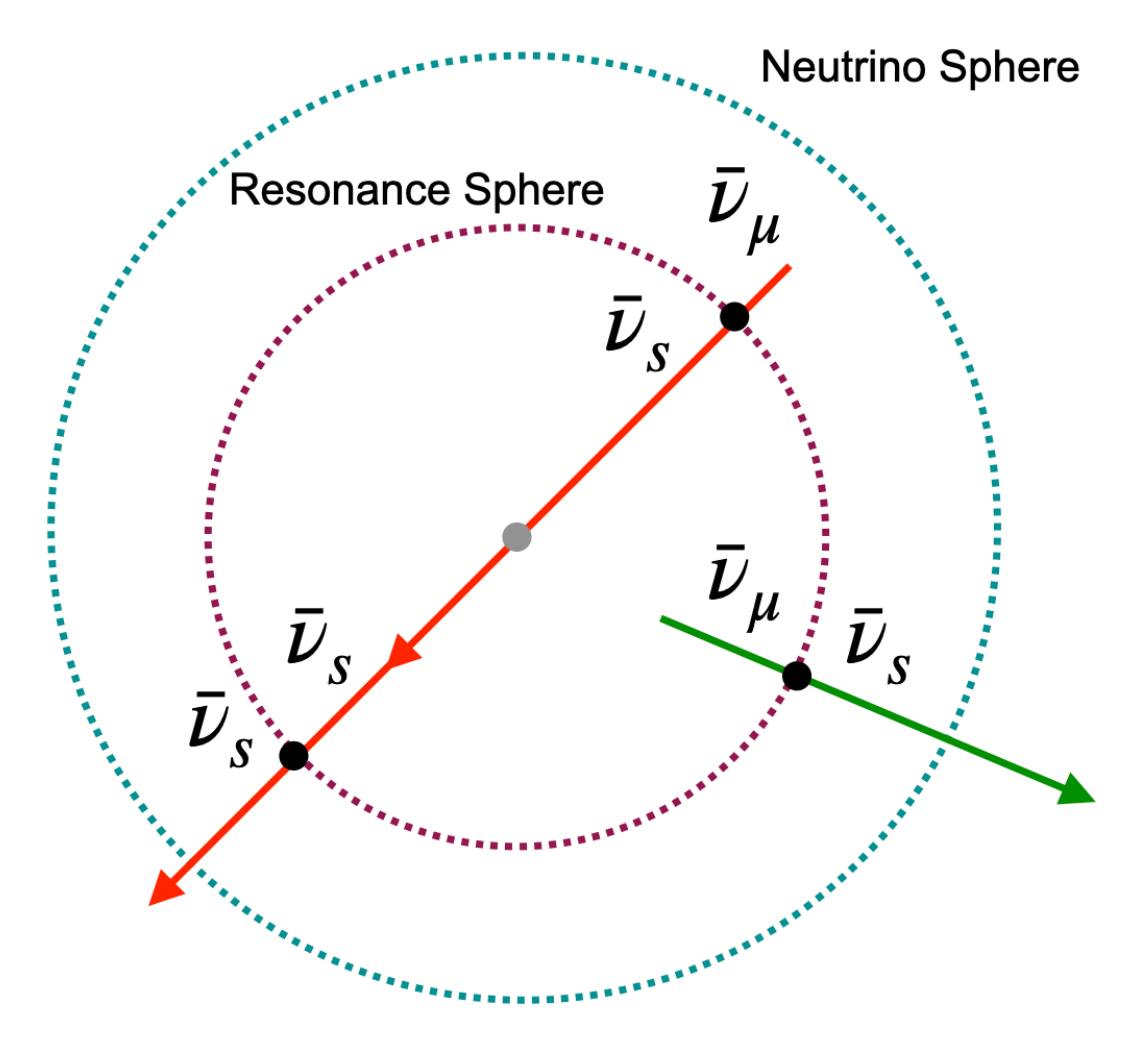}
	\caption{Illustration of two types of escaping $\bar\nu_s$. One type is converted from radially outgoing $\bar\nu_\mu$ that experience
 a single MSW resonance, while the other type is converted from radially incoming $\bar\nu_\mu$ that experience two resonances (with conversion
 occurring only at the first resonance).\label{fig:resonance}}
\end{figure}

Resonant conversion of $\bar{\nu}_{\mu}$ into
$\bar{\nu}_{s}$ is the dominant mechanism for producing sterile neutrinos inside the PNS (e.g., \cite{2019PhRvD..99d3012A,2019JCAP...12..019S,PhysRevD.106.015017,PhysRevD.108.063025}).
We follow Ref.~\cite{PhysRevD.108.063025} and
adopt the following rate of change in $Y_{\nu_\mu}$ due to the MSW conversion of $\bar\nu_\mu$ and the subsequent escape of $\bar\nu_s$
from the PNS:
\begin{align}
	\dot Y_{\nu_\mu}^{\rm MSW}=\Theta(\lambda_R-\delta r)
	\frac{\pi E_R(1-P_{\rm LZ}^2)}{n_bH_R}\left.\frac{d^2n_{\bar\nu_\mu}}{dEd\Omega}\right|_{E_R}.\label{eq:rymsw}
\end{align}
In the above equation, $\Theta(x)$ is the Heaviside step function, $\lambda_R=1/[n_b\sigma(E_R)]$ is the mean free path at the resonant 
neutrino energy $E_R$, $\sigma(E)\approx G_F^2E^2/\pi$ is the cross section governing $\bar\nu_\mu$ diffusion (e.g., \cite{2011PhRvD..83i3014R}),
$\delta r = 2H_R\tan2\theta$ is the width of the resonance region, $H_R$ is the scale height $|\partial\ln V_{\bar\nu}/\partial r|_{E_R}^{-1}$ 
with the derivative taken at the resonance radius for $E_R$, and 
\begin{align}
    P_{\rm LZ}=\exp\left(-\frac{\pi\delta m^2H_R\sin^22\theta}{4E_R\cos2\theta}\right)\,
\end{align}
is the Landau-Zener survival probability for a radially-propagating $\bar{\nu}_{\mu}$ after it crosses the resonance.
Note that a radially-outgoing $\bar{\nu}_{\mu}$ crosses the resonance only once, whereas a radially-incoming $\bar{\nu}_{\mu}$ crosses
the resonance twice (see Fig.~\ref{fig:resonance}). Both situations are taken into account \cite{PhysRevD.108.063025} by Eq.~(\ref{eq:rymsw}).

\begin{figure*}[!t]
\centering
     \hspace*{0.4 cm}\includegraphics[width=0.38\textwidth]{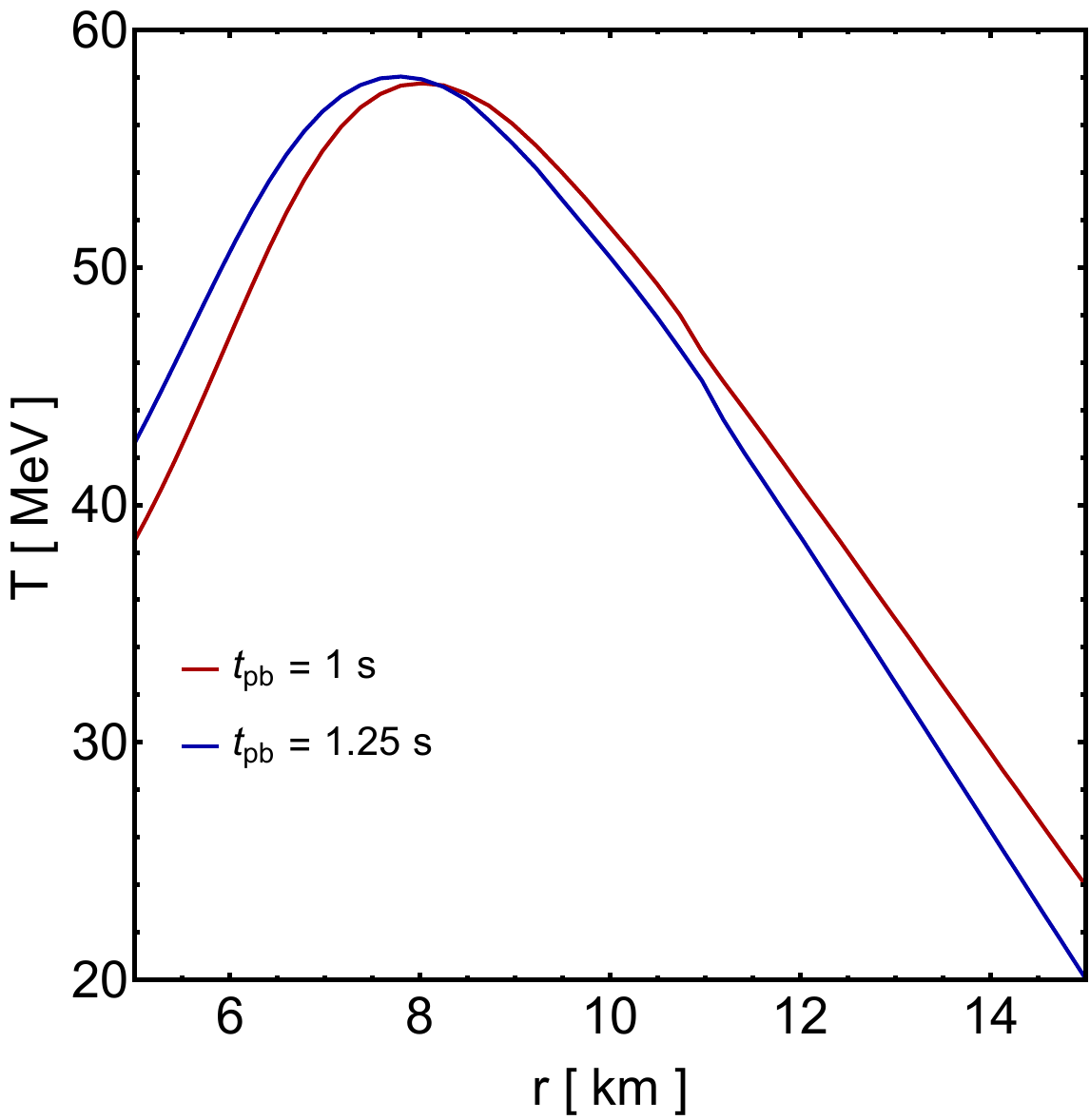}	
     \hspace*{0.4 cm}
      \includegraphics[width=0.425\textwidth]{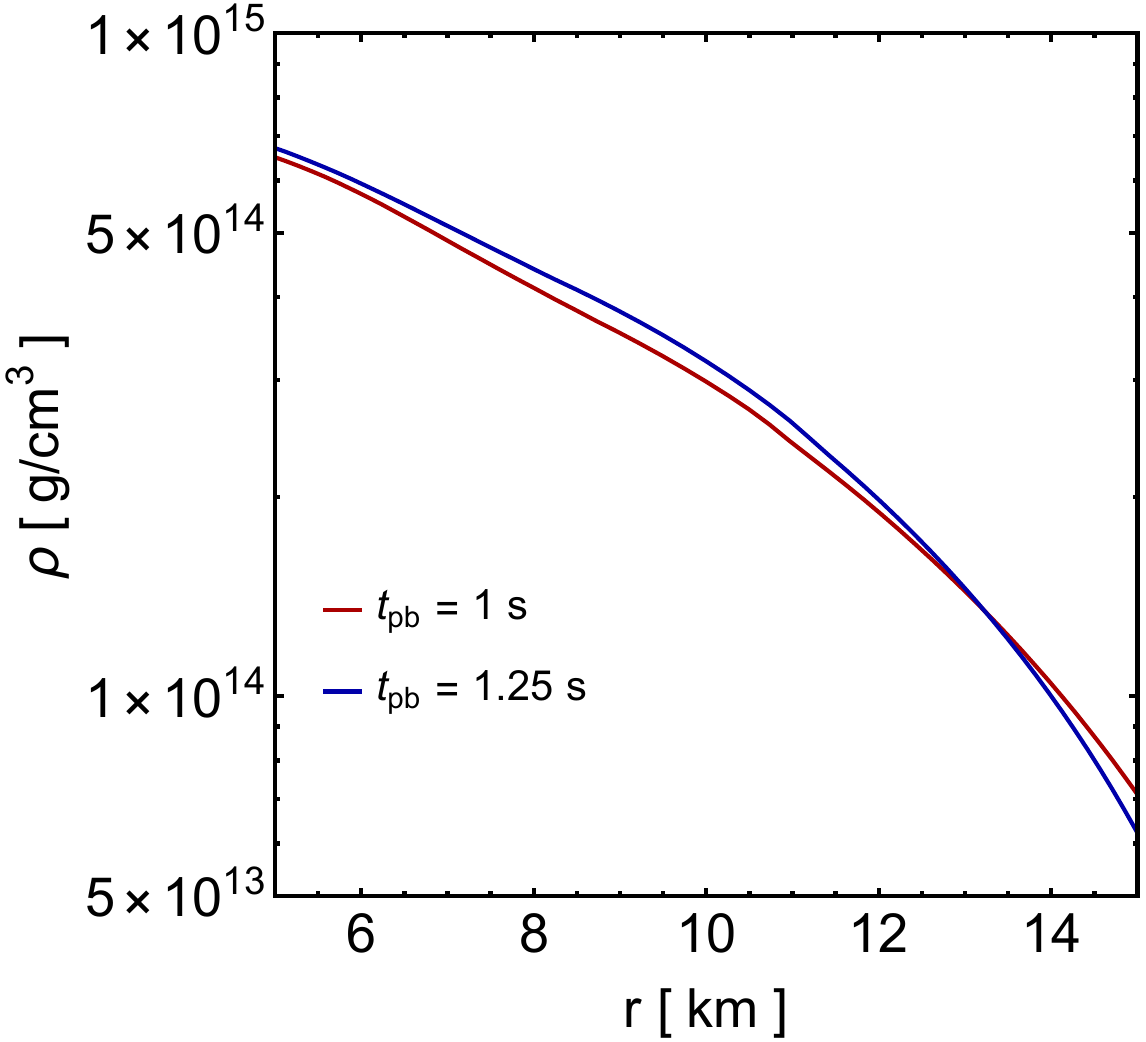}\bigskip\\ 
     \includegraphics[width=0.4\textwidth]{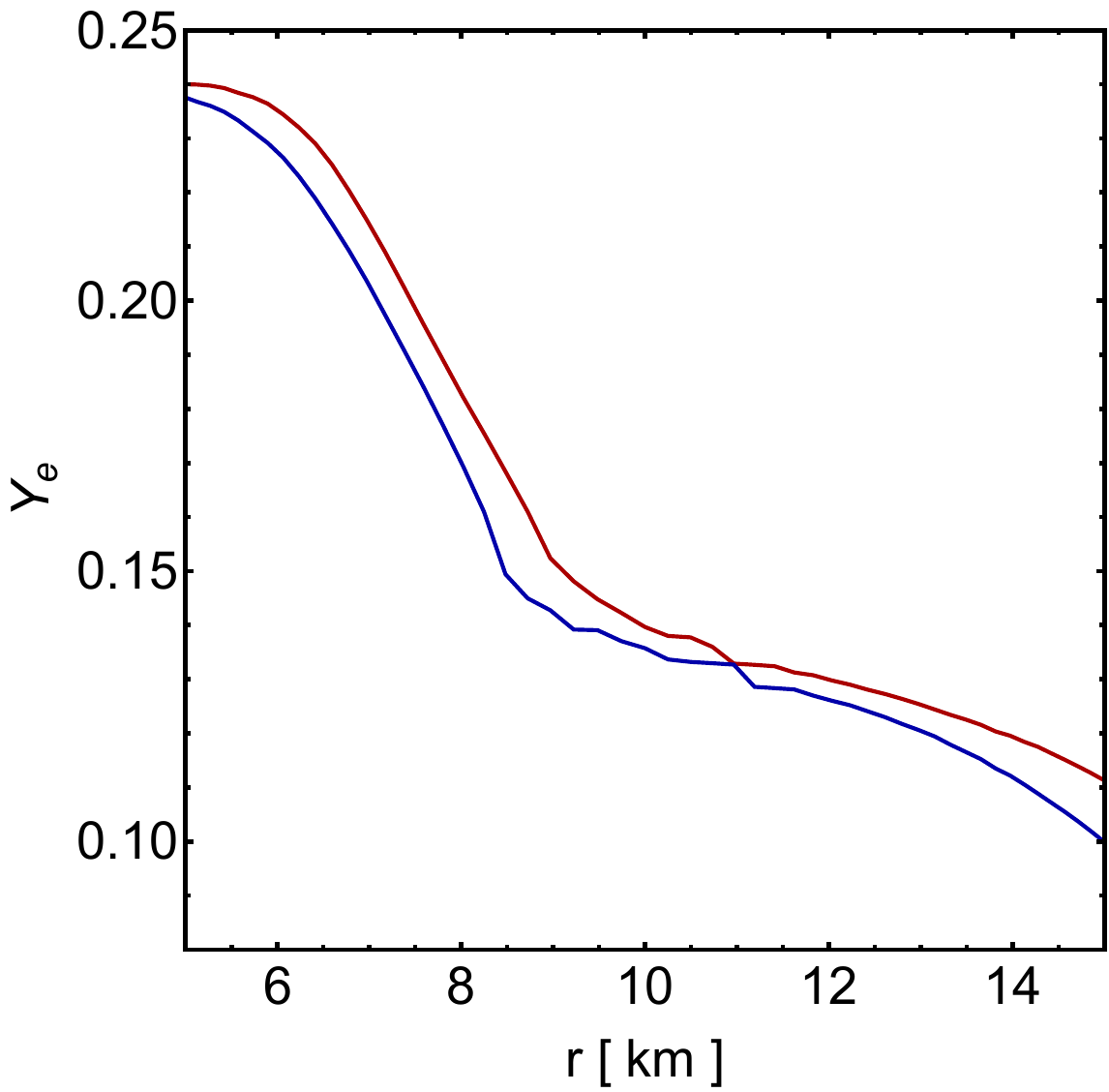}	
     \hspace*{0.8 cm}
     \includegraphics[width=0.4\textwidth]{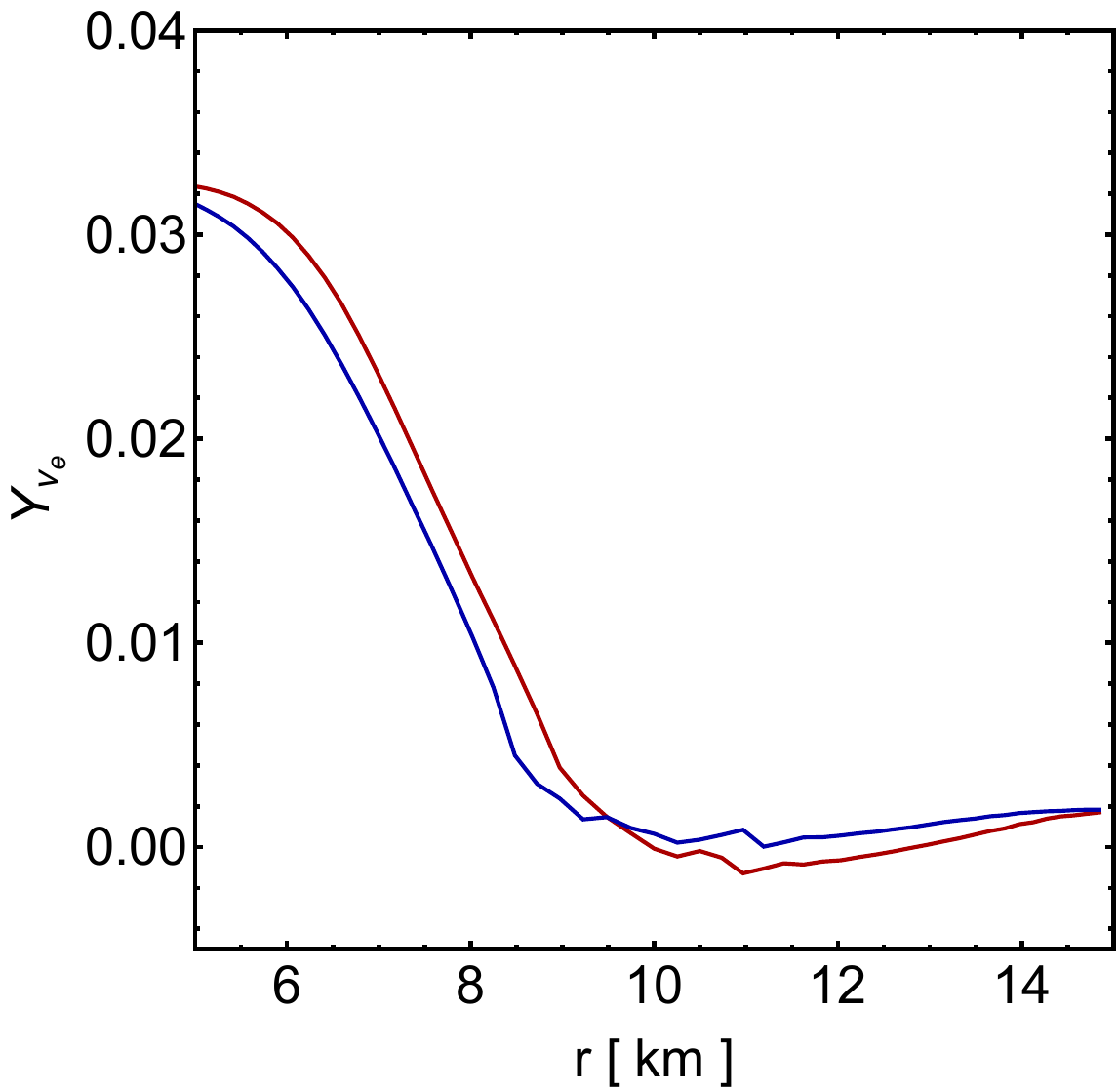}\bigskip\\ 
      \includegraphics[width=0.4\textwidth]{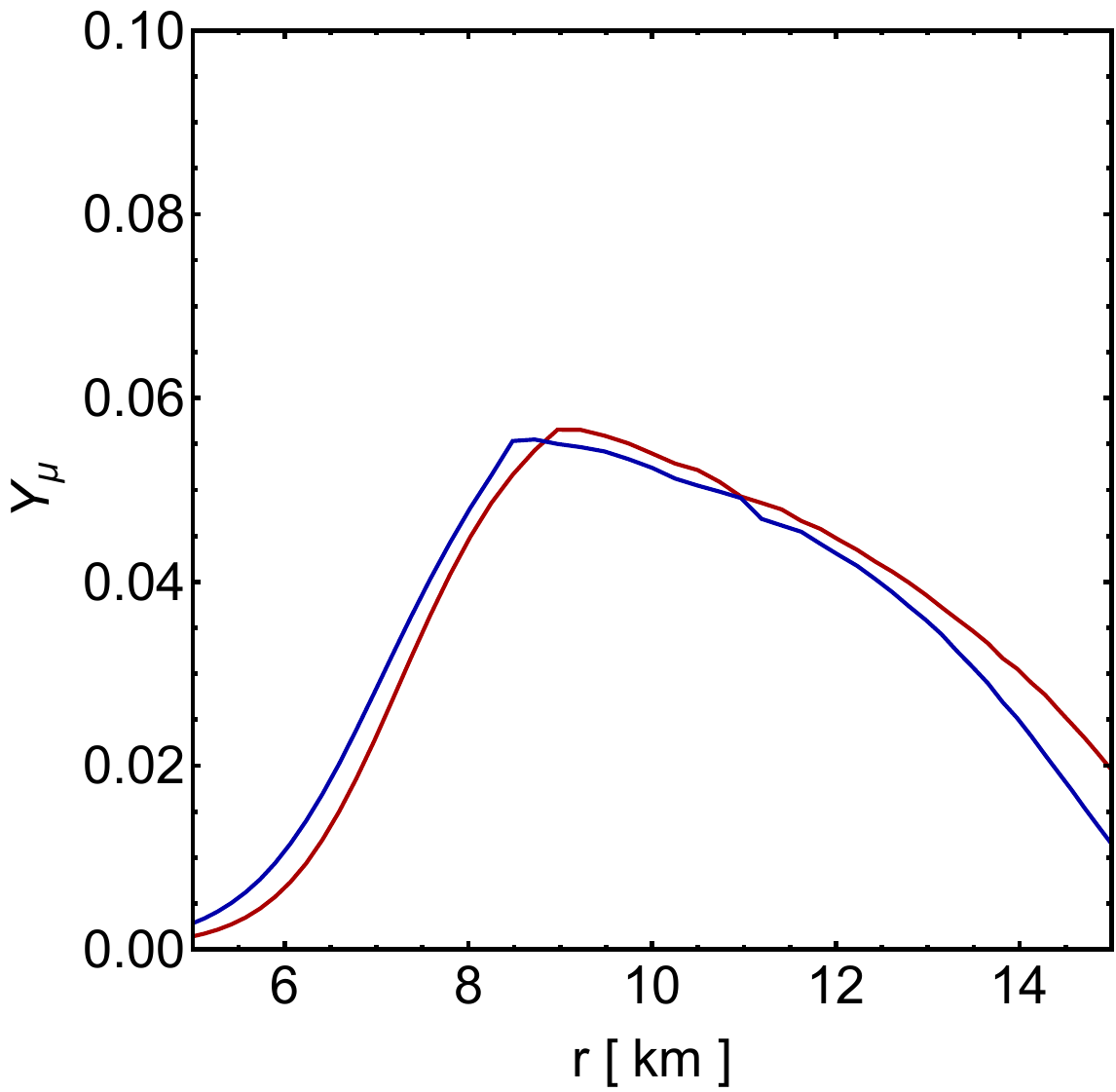}	
    \hspace*{0.5 cm}
     \includegraphics[width=0.42\textwidth]{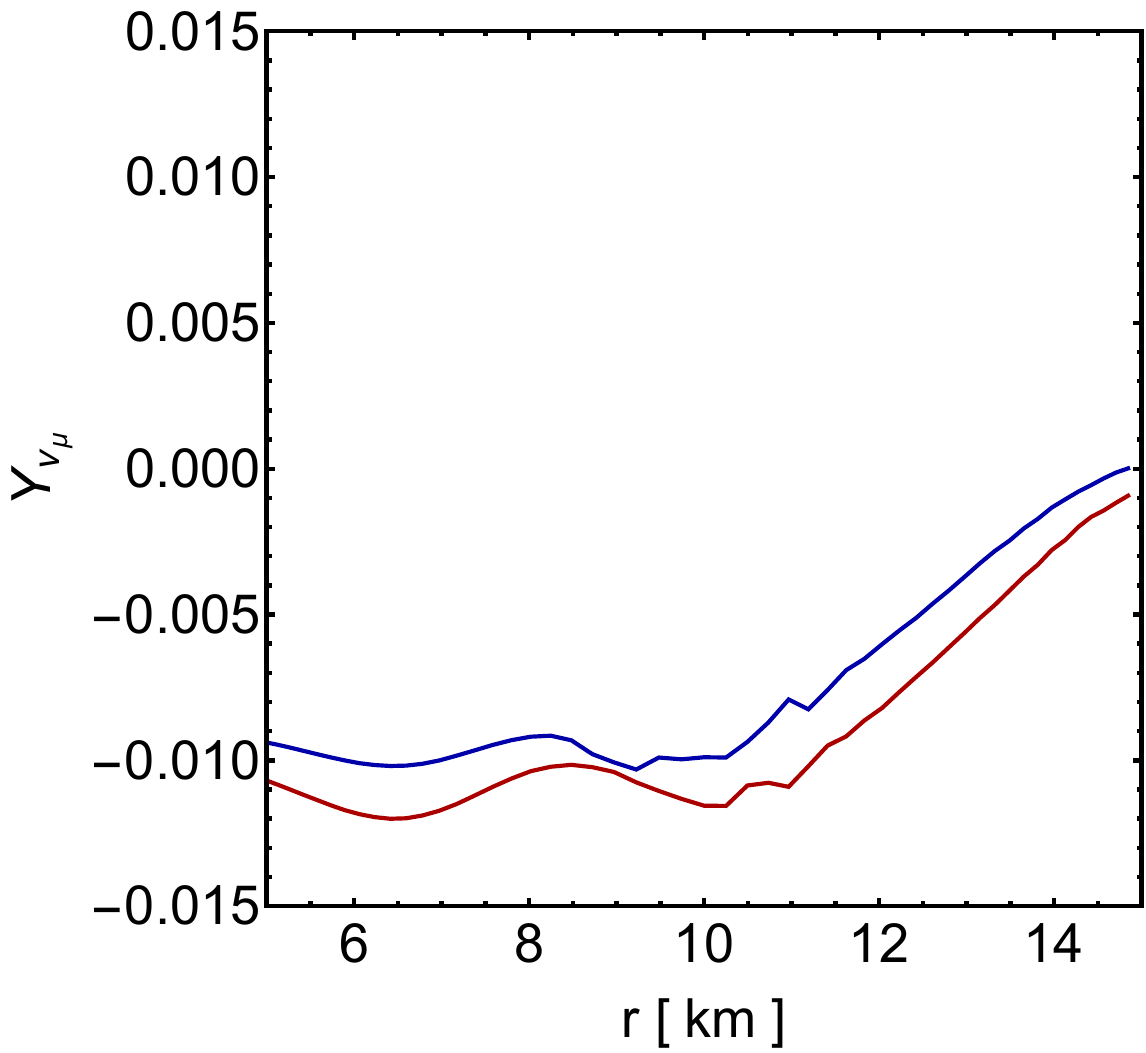}\bigskip\\ 
 \caption{Comparison of the radial profiles of $T$, $\rho$, $Y_e$, $Y_{\nu_e}$, $Y_\mu$, and $Y_{\nu_\mu}$ at
$t_{\rm pb}=1$ and 1.25~s for a $20\,M_{\odot}$ CCSN model (with the SFHo nuclear equation of state and with muonization) provided by the Garching group \cite{2017PhRvL.119x2702B,Bollig2016}.
\label{fig:cond}}
\end{figure*}

Apart from the MSW conversion of $\bar\nu_\mu$
into $\bar\nu_s$, $\nu_{s}$ ($\bar{\nu}_{s}$) can be produced via collisions of $\nu_{\mu}$ ($\bar{\nu}_{\mu}$) with the PNS constituents. 
In this scenario, $\nu_{\mu}$ ($\bar{\nu}_{\mu}$) evolves as a linear combination of two effective mass eigenstates between collisions. Upon collision,
which is predominantly with baryons, the wave function collapses, and a $\nu_s$ ($\bar\nu_s$) is produced with a probability proportional to $\sin^22\theta_\nu$ $(\sin^22\theta_{\bar\nu})$. 
For the mixing parameters of interest, the collisional production of $\nu_s$ and $\bar\nu_s$ inside a PNS is much less efficient than the MSW conversion of $\bar\nu_\mu$
into $\bar\nu_s$. Nevertheless, for completeness, we also include the following contribution to the rate of change in $Y_{\nu_\mu}$ from the collisional production
of $\nu_s$ and $\bar\nu_s$:
	\begin{align}
	\dot Y_{\nu_\mu}^{\rm coll}
	&=G_F^2\left(\int_{E_{\bar\nu_\mu}}dE\,E^2\sin^22\theta_{\bar\nu}
	\frac{d^2n_{\bar\nu_\mu}}{dEd\Omega}\right.\nonumber\\
	&\left.-\int_0^\infty dE\,E^2\sin^22\theta_{\nu}\frac{d^2n_{\nu_\mu}}{dEd\Omega}\right)\,.
\end{align}
In the above equation, the integration over $E_{\bar\nu_\mu}$ excludes the range $[E_R-\delta E/2,E_R+\delta E/2]$ with $\delta E=(\lambda_R/H_R)E_R$ 
when $\lambda_R$ exceeds $\delta r$ and MSW conversion occurs for resonant $\bar\nu_\mu$, but covers the entire energy range otherwise.

\section{Feedback from active-sterile neutrino mixing}
\label{sec:feedback}
We focus on the region where active neutrinos are trapped.
Because sterile neutrinos can escape from this region, $\nu_{\mu}$-$\nu_{s}$ and $\bar\nu_{\mu}$-$\bar\nu_{s}$ mixing
leads to changes of the $\nu_\mu$ and $\bar\nu_\mu$ energy distributions, which in turn cause evolution of $Y_{\nu_\mu}$, 
and hence that of $V_\nu$ and $V_{\bar\nu}$. This type of feedback also occurs for 
$\nu_{\tau}$-$\nu_{s}$ and $\bar\nu_{\tau}$-$\bar\nu_{s}$ mixing \cite{2011PhRvD..83i3014R,2019JCAP...12..019S,PhysRevD.106.015017,PhysRevD.108.063025}. 
What is new here, however, is that the changes of
the $\nu_\mu$ and $\bar\nu_\mu$ energy distributions cause evolution of $Y_n$, $Y_p$, and $Y_\mu$ through the reactions 
\begin{subequations}
\begin{align}
    \nu_\mu+n&\to p+\mu^-,\label{eq:numun}\\
    \bar\nu_\mu+p&\to n+\mu^+.\label{eq:bnumup}
\end{align}
\end{subequations}
Therefore, the feedback effects for $\nu_{\mu}$-$\nu_{s}$ and $\bar\nu_{\mu}$-$\bar\nu_{s}$ mixing are more complicated
than those for $\nu_{\tau}$-$\nu_{s}$ and $\bar\nu_{\tau}$-$\bar\nu_{s}$ mixing.
Reactions in Eqs.~(\ref{eq:numun}) and (\ref{eq:bnumup}) also occur in reverse, and there are parallel processes
involving $\nu_e$ and $\bar\nu_e$ as well:
\begin{subequations}
\begin{align}
    \nu_e+n&\rightleftharpoons p+e^-,\label{eq:nuen}\\
    \bar\nu_e+p&\rightleftharpoons n+e^+.\label{eq:bnuep}
\end{align}
\end{subequations}
Consequently, in order to treat the feedback on $V_\nu$ and $V_{\bar\nu}$ for $\nu_{\mu}$-$\nu_{s}$ and $\bar\nu_{\mu}$-$\bar\nu_{s}$ mixing,
we must include all the processes described above to follow the evolution of $Y_n$, $Y_p$, $Y_\mu$, $Y_e$, $Y_{\nu_\mu}$, and $Y_{\nu_e}$ simultaneously. 

As mentioned in Section~\ref{sec:setup}, the particles of concern, $n$, $p$, $\mu^\pm$, $e^\pm$, $\nu_\mu$, $\bar\nu_\mu$, $\nu_e$, and $\bar\nu_e$, 
are assumed to have Fermi-Dirac energy distributions characterized by the same temperature $T$ but species-specific chemical potentials.
For fixed $T$ and $n_b$, the number fractions $Y_n$, $Y_p$, $Y_\mu$, $Y_e$, $Y_{\nu_\mu}$, and $Y_{\nu_e}$ are specified by the corresponding
chemical potentials [see e.g., Eq.~(\ref{eq:ynumu}) for $Y_{\nu_\mu}$]. We use the following six constraints to determine these chemical potentials:
\begin{subequations}
\begin{align}
       Y_n+Y_p&=1,\label{eq:ynp}\\
       Y_e+Y_\mu&=Y_p,\label{eq:yemup}\\
       \mu_{\nu_e}+\mu_n&=\mu_p+\mu_e,\label{eq:munuenpe}\\
  {\mu_{\nu_\mu}+\mu_n}&{=\mu_p+\mu_\mu},\label{eq:munumunpmu}\\
       \dot{Y}_e+\dot{Y}_{\nu_e}&=\Gamma_{L_e},\label{eq:le}\\
       \dot{Y}_{\mu}+\dot{Y}_{\nu_{\mu}}&=\Gamma_{L_\mu}\label{eq:lmu},
\end{align}
\end{subequations}
where $\Gamma_{L_e}$ and $\Gamma_{L_\mu}$ are the net rates of change in $Y_e+Y_{\nu_e}$ and $Y_\mu+Y_{\nu_\mu}$ (see below).
The constraints in Eqs.~(\ref{eq:ynp}), (\ref{eq:yemup}), (\ref{eq:le}), and (\ref{eq:lmu}) correspond to conservation of baryon number, electric charge, 
and net lepton numbers of the electron and muon types. We have assumed that chemical equilibrium is achieved
among $n$, $p$, $e^\pm$, $\nu_e$, and $\bar\nu_e$ [Eq.~(\ref{eq:munuenpe})], and also among 
$n$, $p$, $\mu^\pm$, $\nu_\mu$, and $\bar\nu_\mu$ [Eq.~(\ref{eq:munumunpmu})]. This assumption is consistent with the Fermi-Dirac energy distributions
adopted for the above particles and with the fast rates (see Appendix) for interconverting $n$ and $p$ by these particles.

For the net rates of change in $Y_e+Y_{\nu_e}$ and $Y_\mu+Y_{\nu_\mu}$, we ignore the transport of $Y_{\nu_e}$ and $Y_{\nu_\mu}$ 
by diffusion and take 
\begin{subequations}
\begin{align}
    \Gamma_{L_e}&=0,\label{eq:gammaleno}\\
    \Gamma_{L_\mu}&=\dot Y_{\nu_\mu}^{\rm MSW}+\dot Y_{\nu_\mu}^{\rm coll}.\label{eq:gammalmuno}
\end{align}
\end{subequations}
The above approximation highlights the effects of sterile neutrino production and its validity will be discussed in Section~\ref{sec:discuss}.

\begin{figure*}[!t]
\centering
\includegraphics[width=0.41\textwidth]{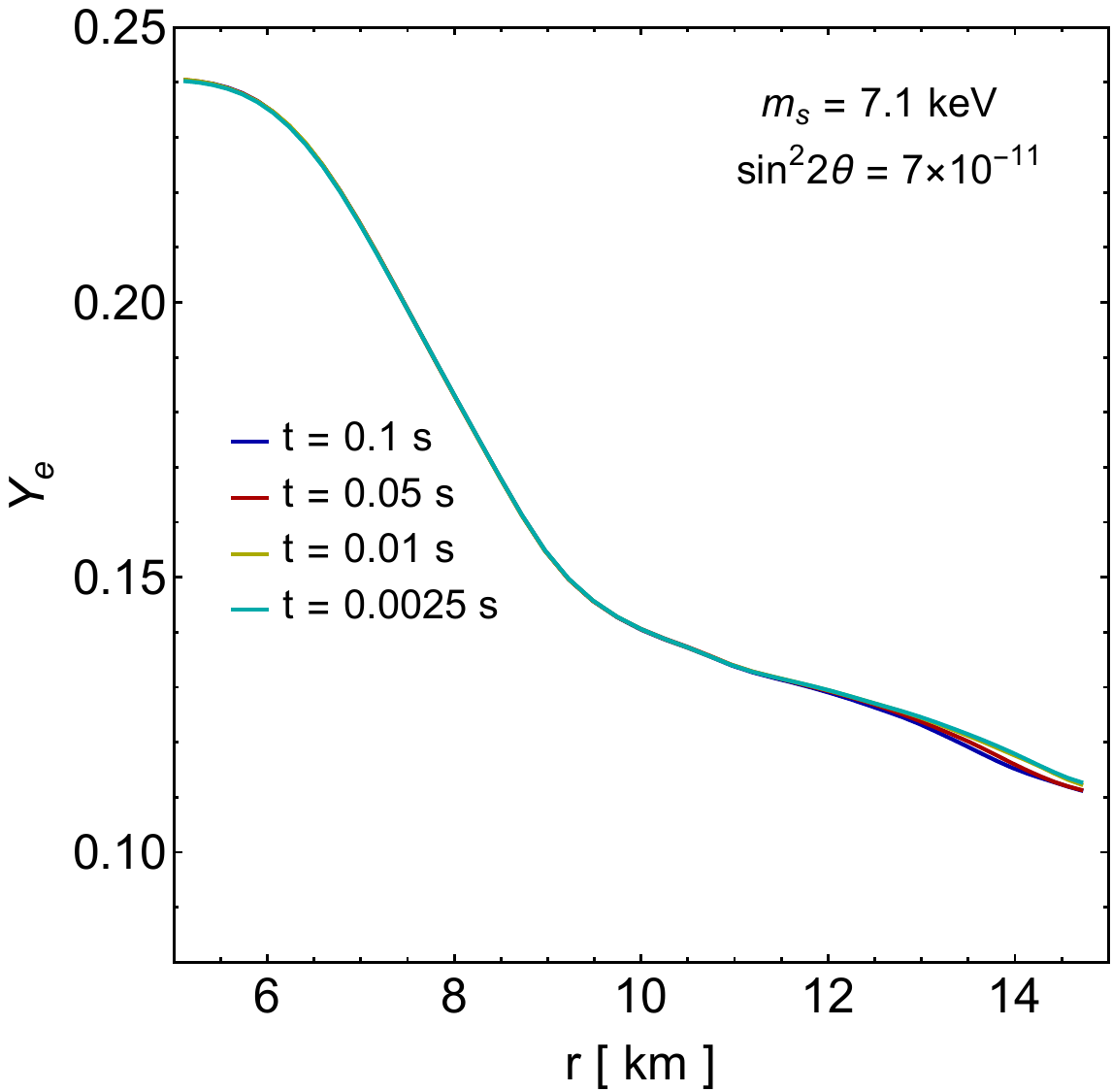}	
\hspace*{0.75 cm}
\includegraphics[width=0.425\textwidth]{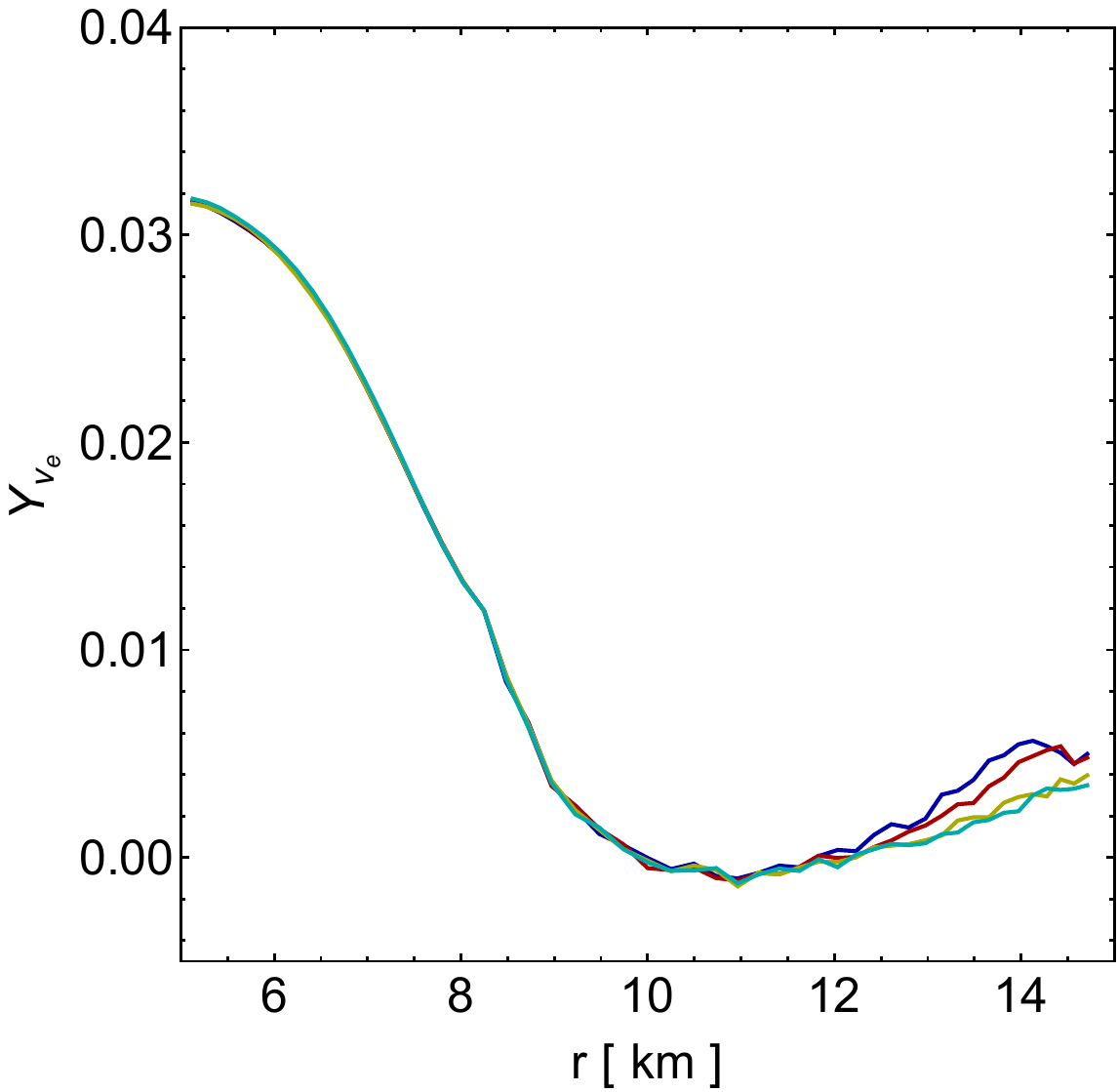}\bigskip\\	
\includegraphics[width=0.41\textwidth]{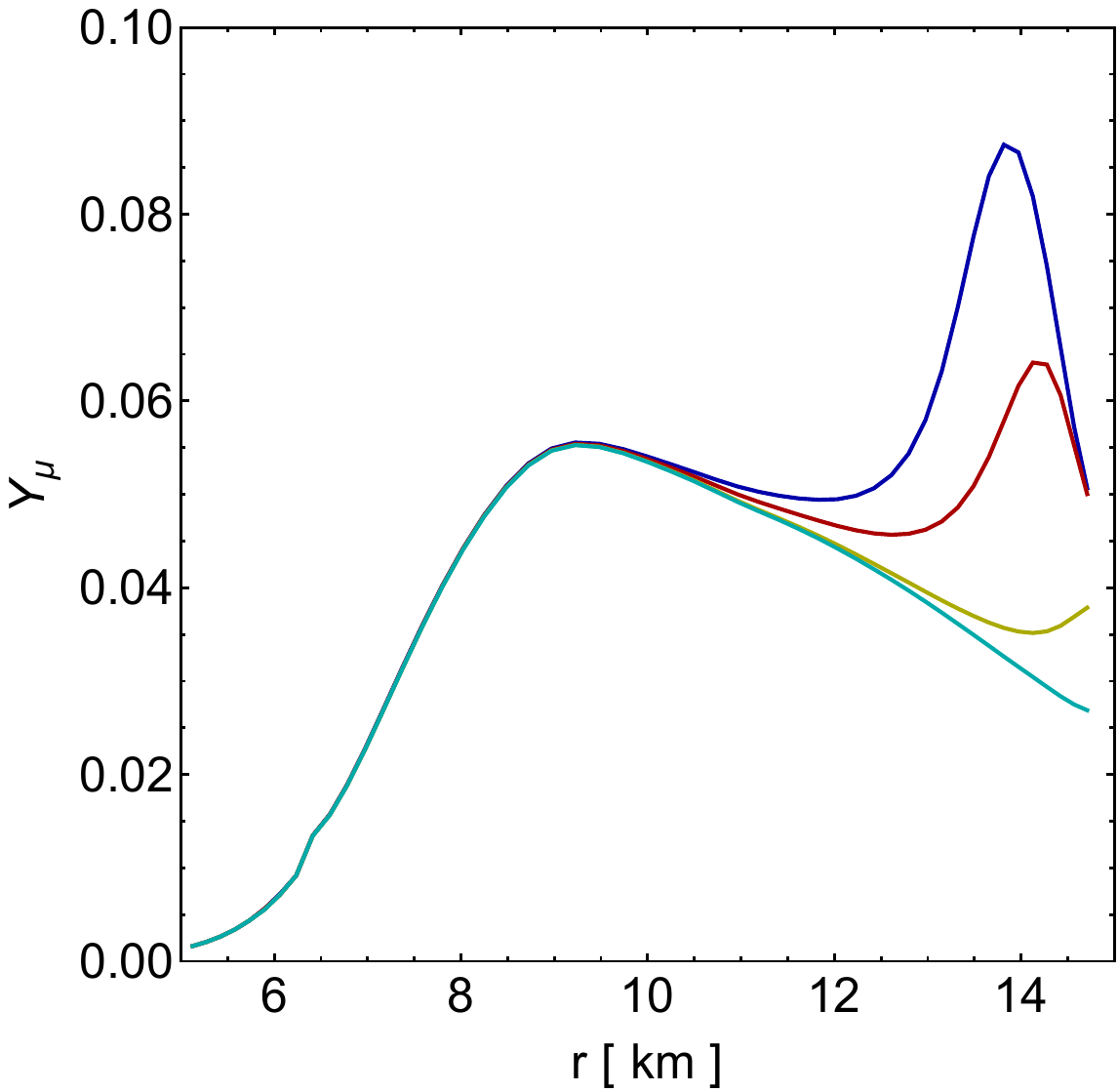}	
\hspace*{0.75 cm}
\includegraphics[width=0.425\textwidth]{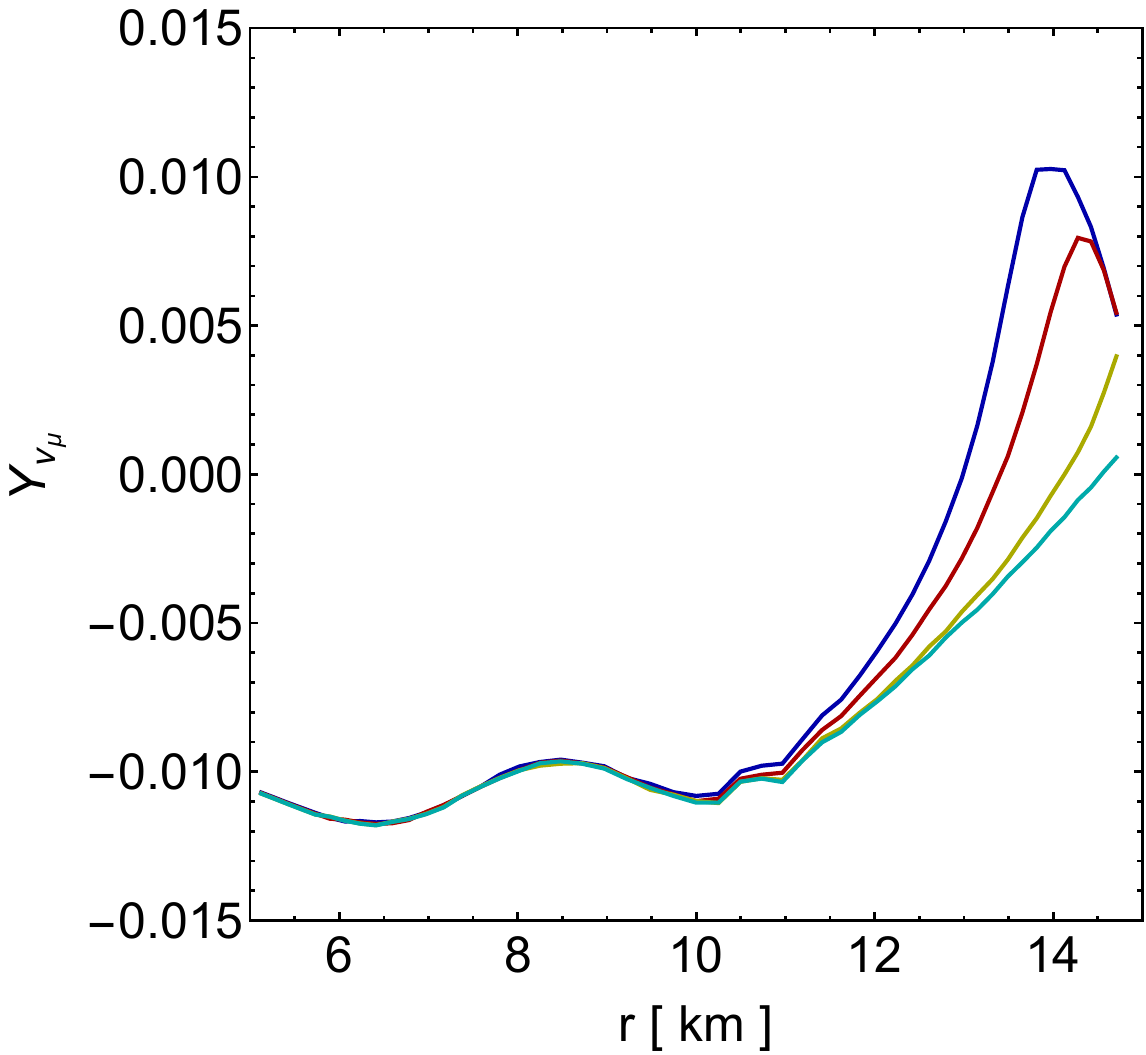}	
\caption{Evolution of $Y_e$, $Y_{\nu_e}$, $Y_\mu$, and $Y_{\nu_\mu}$ as functions of
radius $r$ and time $t$ for $m_s=7.1$~keV and $\sin^22\theta=7\times 10^{-11}$.
Diffusion of $Y_{\nu_e}$ and $Y_{\nu_\mu}$ is ignored.\label{fig:yevol7.1}}
\end{figure*}

\begin{figure*}[!t]
\centering
\includegraphics[width=0.41\textwidth]{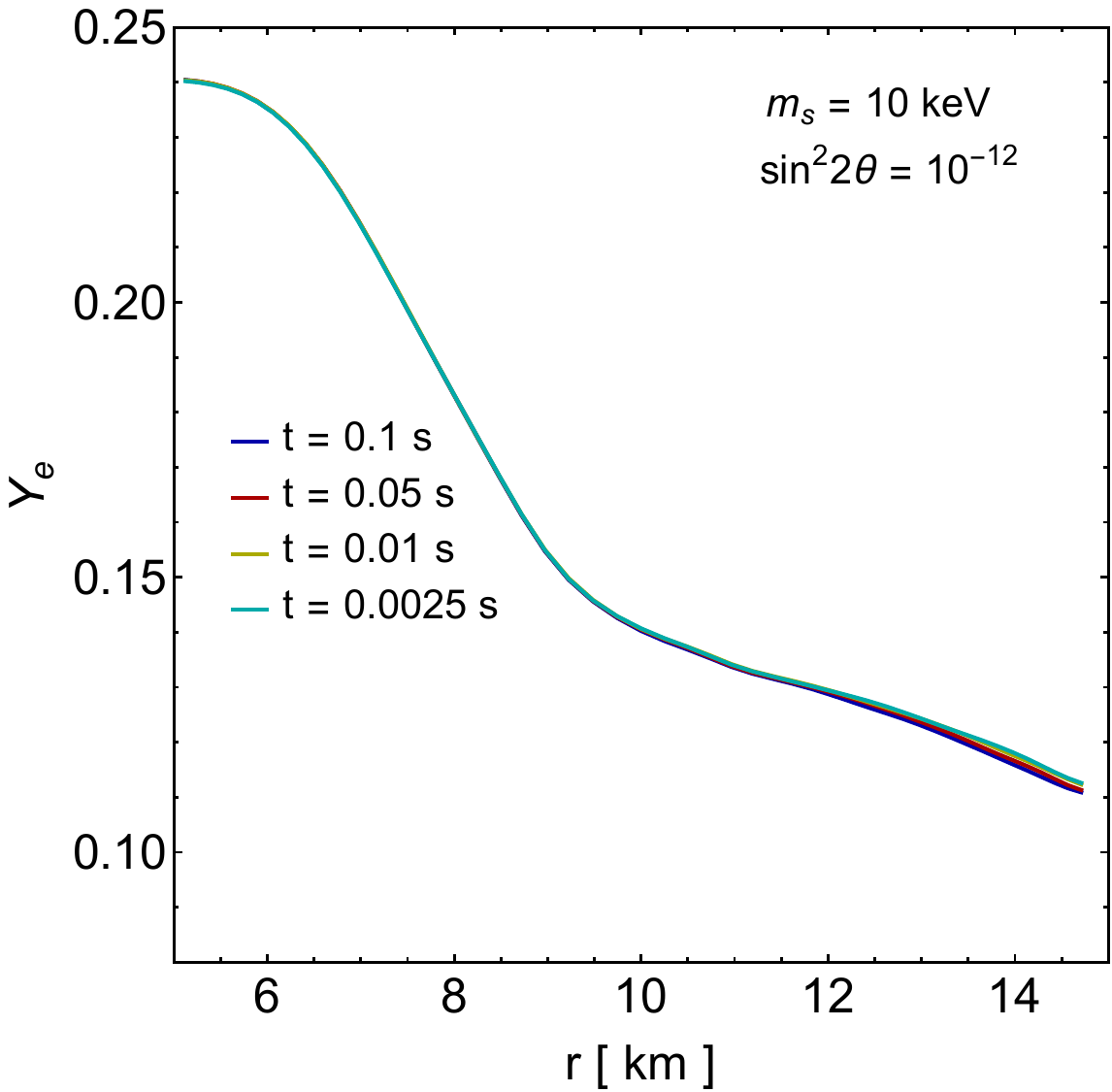}	
\hspace*{0.75 cm}
\includegraphics[width=0.425\textwidth]{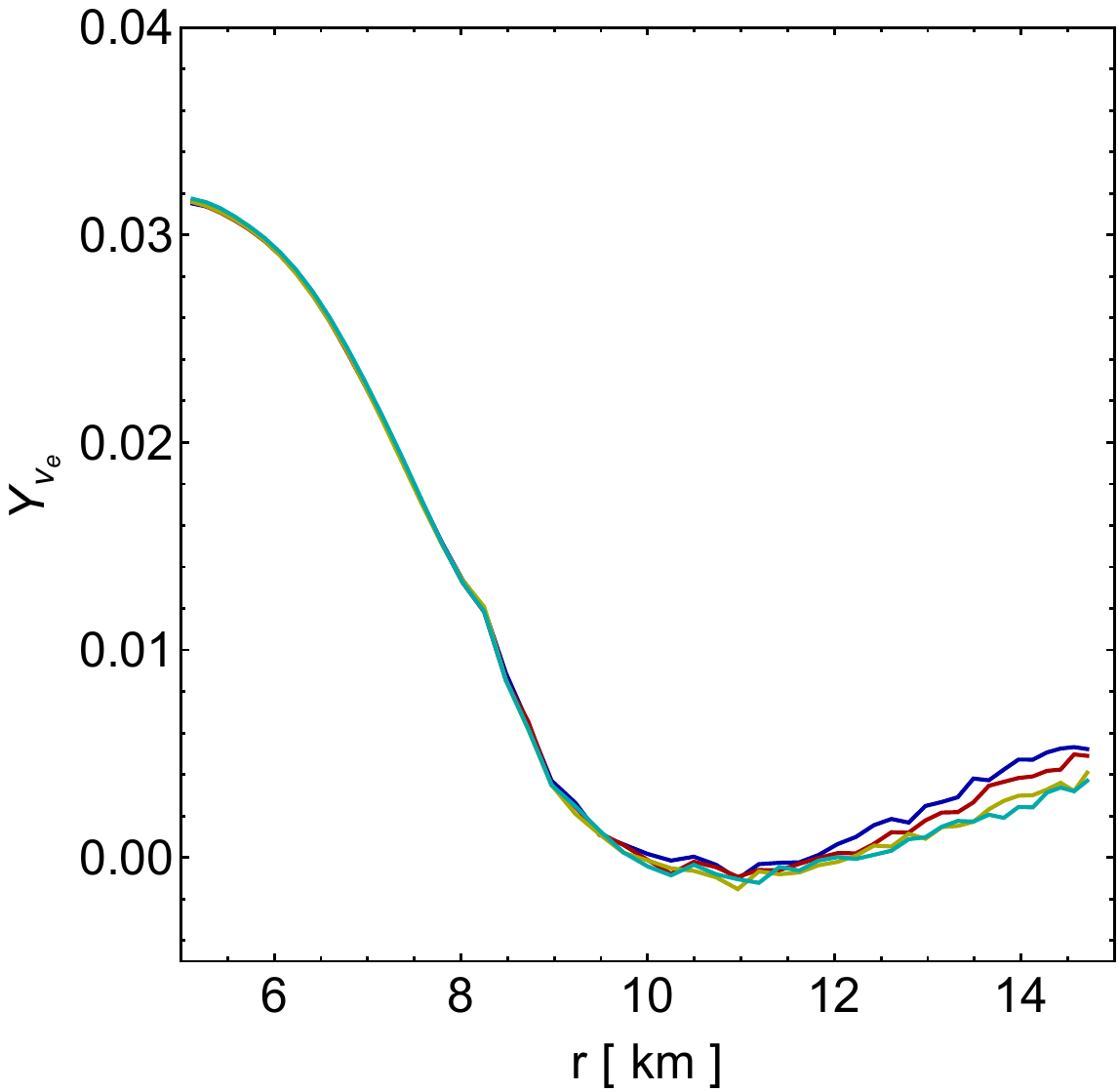}\bigskip\\	
\includegraphics[width=0.41\textwidth]{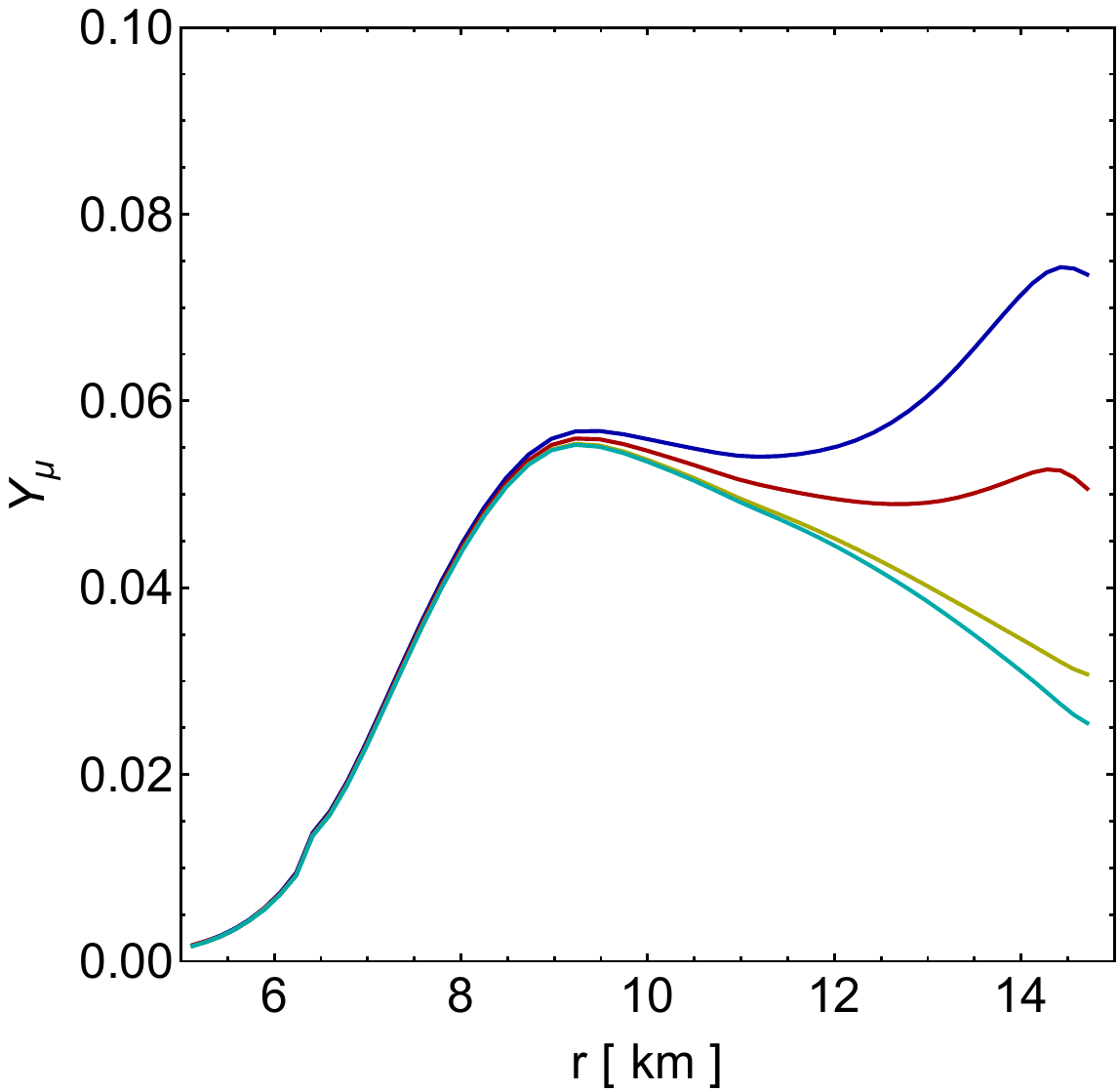}	
\hspace*{0.75 cm}
\includegraphics[width=0.425\textwidth]{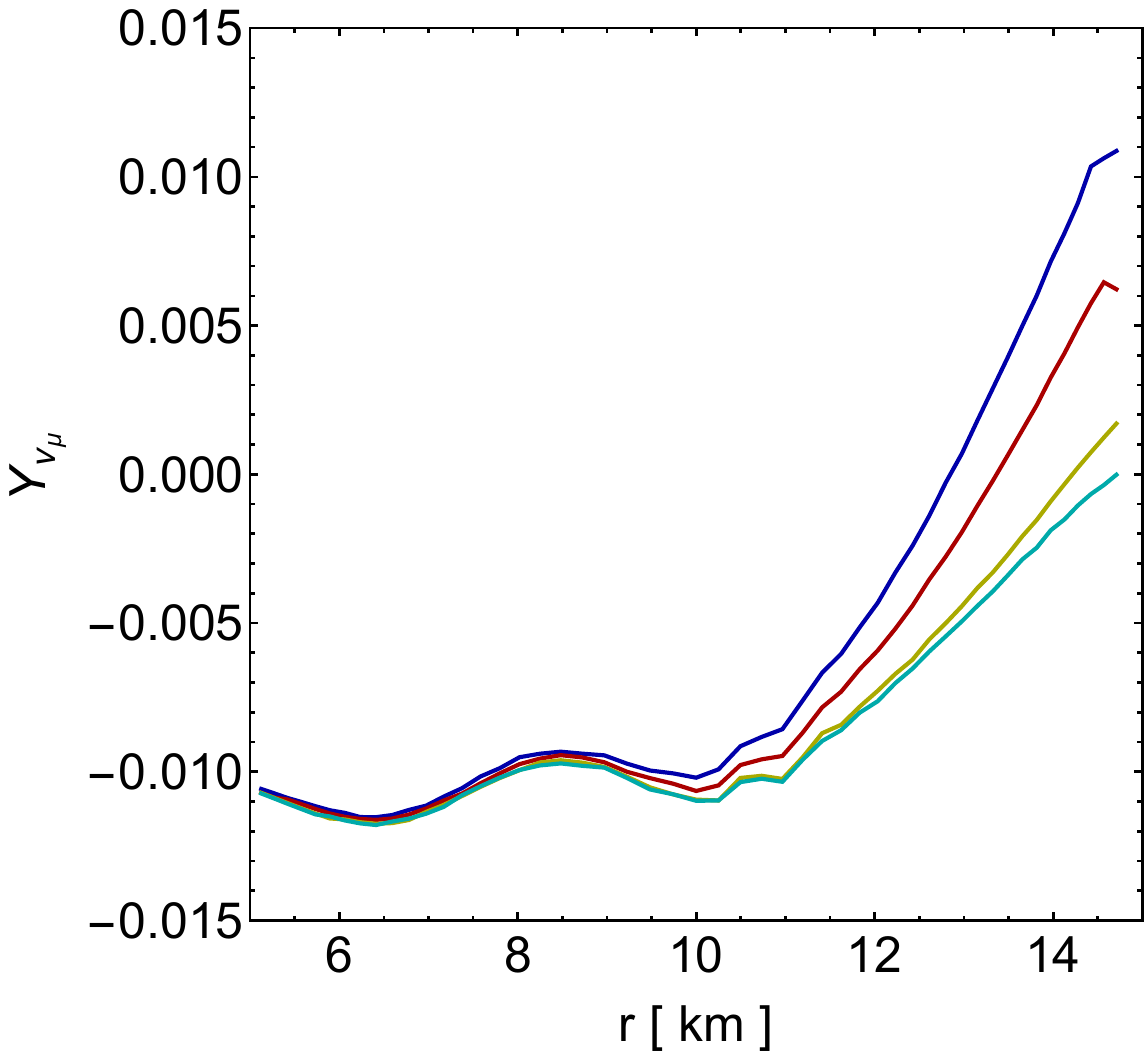}	
\caption{Similar to Fig.~\ref{fig:yevol7.1}, but for $m_s=10$~keV and $\sin^22\theta=10^{-12}$.\label{fig:yevol10}}
\end{figure*}

\section{Results}
\label{sec:example}
We now present example calculations of the effects of $\nu_{\mu}$-$\nu_{s}$ and $\bar\nu_{\mu}$-$\bar\nu_{s}$ mixing in a PNS.
We expect that in addition to the feedback effects discussed in Section~\ref{sec:feedback}, such mixing would also affect
the evolution of temperature $T$ and density $\rho$. As mentioned in the introduction, such dynamic effects can be properly 
quantified only by including $\nu_{\mu}$-$\nu_{s}$ and $\bar\nu_{\mu}$-$\bar\nu_{s}$ mixing in CCSN simulations. Our goal here, 
however, is to elucidate how such mixing enhances muonization. As a reasonable approximation, we limit the duration of our
calculations so that the radial profiles of $T$ and $\rho$ can be considered as fixed in time. As the initial conditions,
we take a snapshot at time post bounce $t_{\rm pb}=1$~s of a $20\,M_{\odot}$ CCSN model (with the SFHo nuclear
equation of state and with muonization) provided by the Garching group \cite{2017PhRvL.119x2702B,Bollig2016}.
Based on the comparison of the conditions at $t_{\rm pb}=1$~s and those at the next output instant $t_{\rm pb}=1.25$~s
(see Fig.~\ref{fig:cond}), we consider it reasonable to assume no evolution of $T(r)$ and $\rho(r)$ for a duration of $\Delta t=0.1$~s.
We also limit our calculations to within the radius $r=14.7$~km, which corresponds to a temperature of $T=25.6$~MeV and
a density of $\rho=8.01\times 10^{13}$~g~cm$^{-3}$,
so that all active neutrinos are in the trapped regime throughout the calculations.

Starting with the conditions at $t_{\rm pb}=1$~s and taking $T(r)$ and $\rho(r)$ as fixed in time, we follow the evolution of
$Y_n$, $Y_p$, $Y_e$, $Y_{\nu_e}$, $Y_\mu$, and $Y_{\nu_\mu}$ by solving Eqs.~(\ref{eq:ynp})--(\ref{eq:lmu}) simultaneously. 
We carry out the calculations up to $t_{\rm pb}=1.1$~s in multiple time steps. For each step, we require that the results have numerically converged 
(with differences less than 1\% when the calculations are repeated with half the step size). We have also checked that throughout 
the calculations, the changes in the total energy density and pressure of each radial zone are small 
(always less than 10\% but typically few percent or less) as compared to the initial values of these quantities.
For illustration, we show the results for $t_{\rm pb}=1.0025$, 1.01, 1.05, and 1.1~s.

We take $\delta m^2\approx m_s^2$ with $m_s=7.1$~keV and $\sin^22\theta=7\times 10^{-11}$,
which are suggested by interpreting the X-ray line emission near 3.55 keV from galaxy clusters as due to sterile neutrino 
decay \cite{Bulbul:2014sua,Boyarsky:2014jta}. For comparison, we use $m_s=10$~keV and $\sin^22\theta=10^{-12}$,
which are chosen for sterile neutrinos to make up all the dark matter \cite{Ng:2019gch,Roach:2022lgo}.
For convenience, hereafter we refer to the two adopted sets of mixing parameters by the corresponding $m_s$ values. 
Using the rates $\Gamma_{L_e}$ and $\Gamma_{L_\mu}$ in Eqs.~(\ref{eq:gammaleno}) and (\ref{eq:gammalmuno}), 
we show the evolution of $Y_e$ (top left), $Y_{\nu_e}$ (top right), $Y_\mu$ (bottom left), and $Y_{\nu_\mu}$ (bottom right) 
as functions of radius $r$ and time $t$ in Figs.~\ref{fig:yevol7.1} and \ref{fig:yevol10} for $m_s=7.1$ and 10~keV, respectively.
The evolution of $Y_n$ and $Y_p$ (not shown)
can be obtained from that of $Y_e$ and $Y_\mu$ through Eqs.~(\ref{eq:ynp}) and (\ref{eq:yemup}).

\begin{figure*}[!t]
\centering
\includegraphics[width=0.4\textwidth]{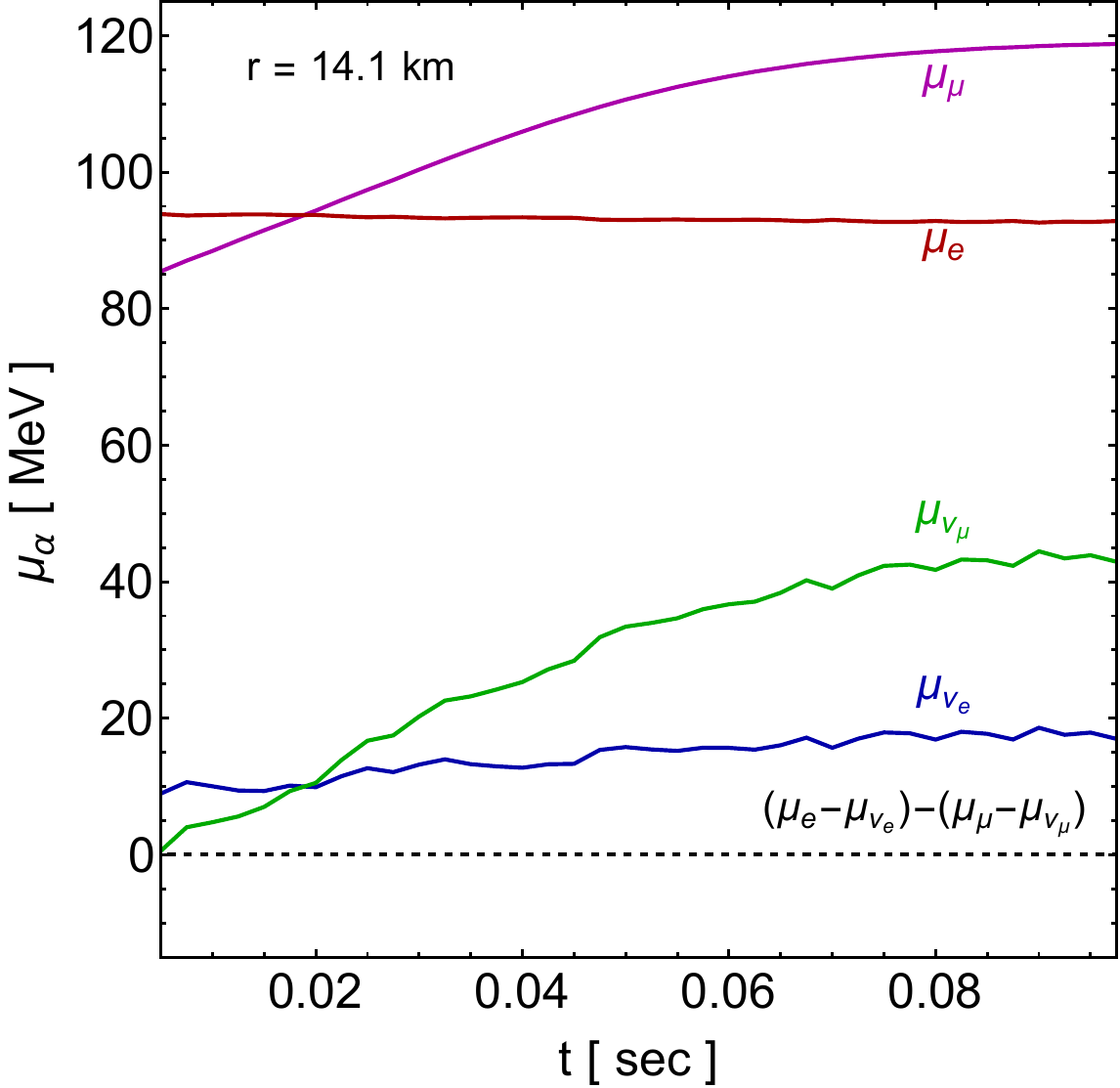}	
\hspace*{0.7 cm}
\includegraphics[width=0.4\textwidth]{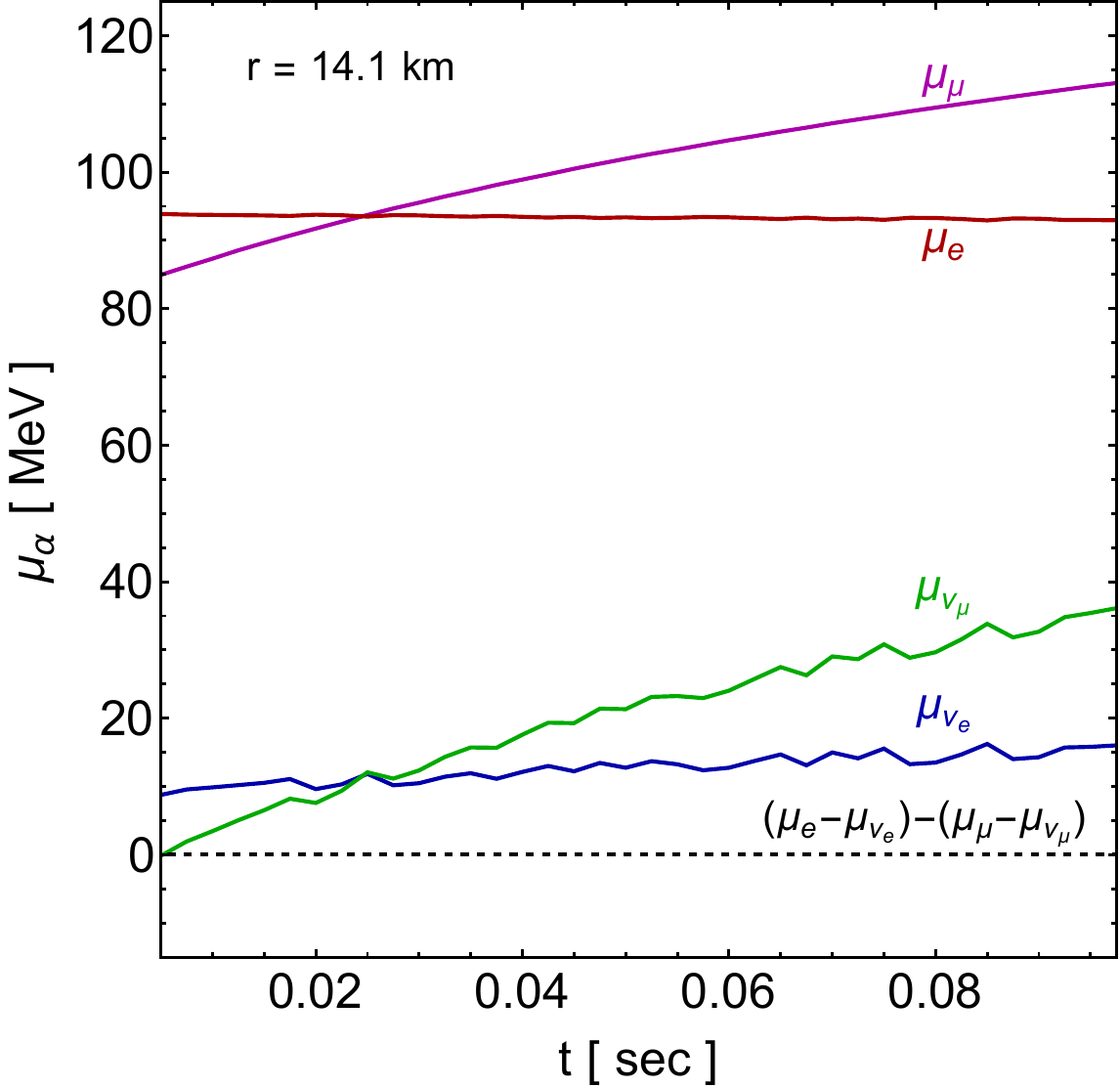}
\caption{Evolution of the chemical potentials $\mu_e$, $\mu_{\nu_e}$, $\mu_\mu$, and $\mu_{\nu_\mu}$ 
as functions of time $t$ for the zone at $r=14.1$~km. The left (right) panel shows the results for $m_s=7.1\,(10)$~keV and $\sin^22\theta=7\times10^{-11}\,(10^{-12})$.
Chemical equilibrium among the relevant particles gives $(\mu_e-\mu_{\nu_e})-(\mu_\mu-\mu_{\nu_\mu})=0$.
\label{fig:chemev}}
\end{figure*}

\begin{figure*}[!t]
\centering
\includegraphics[width=0.4\textwidth]{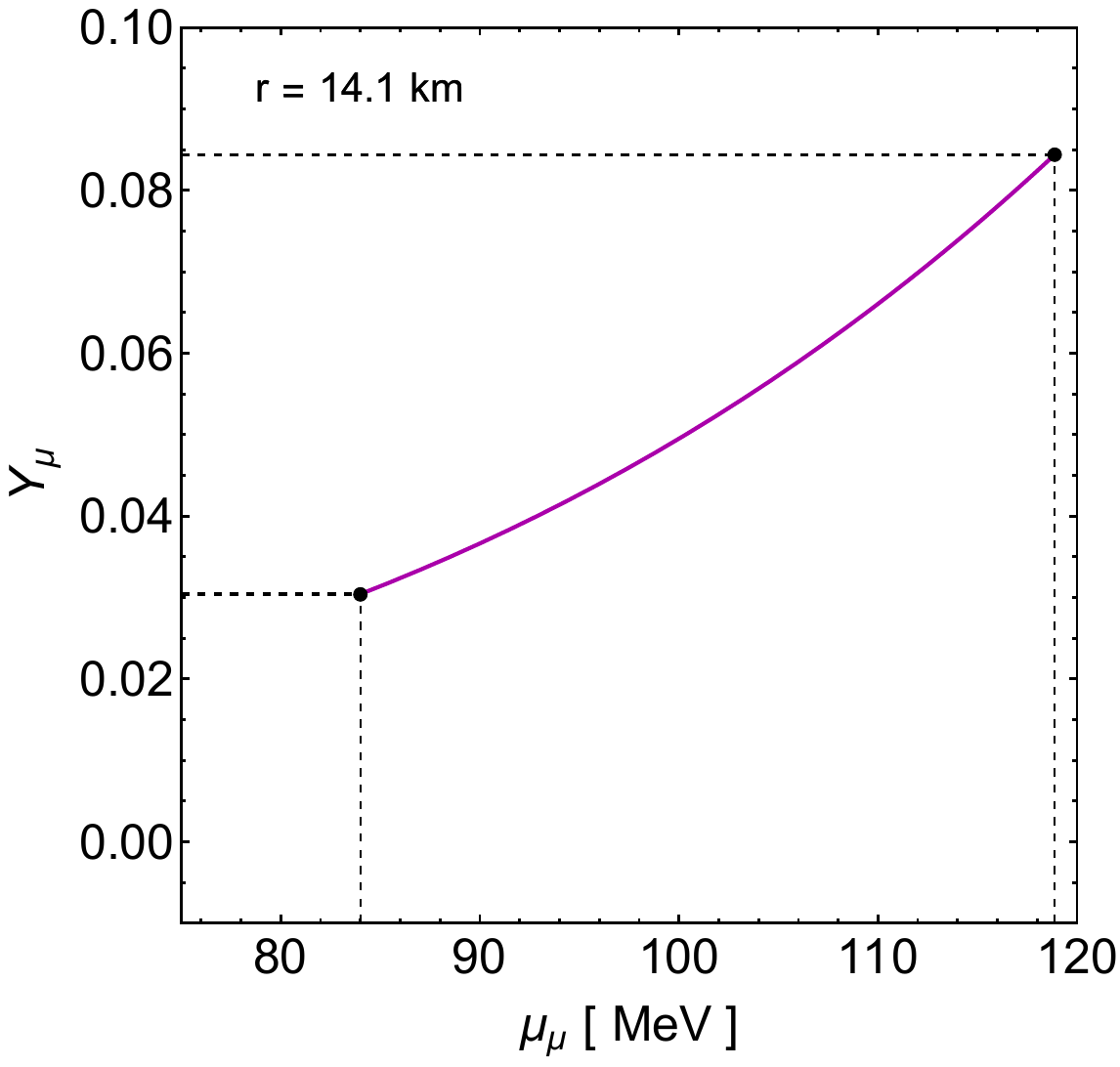}	
\hspace*{0.7 cm}
\includegraphics[width=0.4\textwidth]{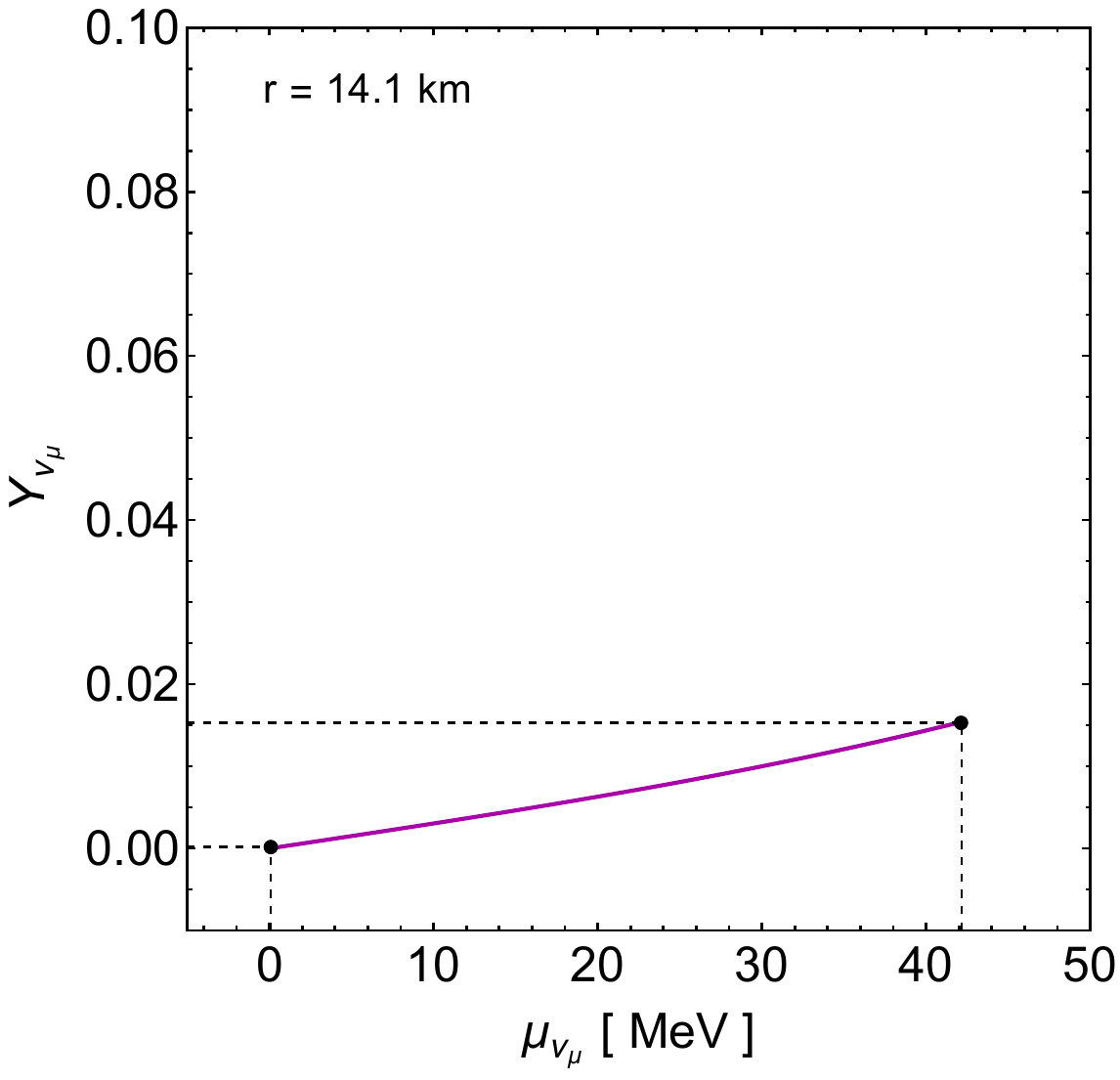}
\caption{Net number fractions $Y_\mu$ (left panel) and $Y_{\nu_\mu}$ (right panel) as functions of 
chemical potentials $\mu_\mu$ and $\mu_{\nu_\mu}$, respectively, for the zone at $r=14.1$~km with
$T=28.9$~MeV and $\rho=9.99\times10^{13}$~g~cm$^{-3}$. The solid points
indicate the values at $t_{\rm pb}=1$ and 1.1~s from the calculations
without diffusion of $Y_{\nu_e}$ and $Y_{\nu_\mu}$ 
for $m_s=7.1$~keV and $\sin^22\theta=7\times10^{-11}$. Because of the muon rest mass,
the increase of $Y_\mu$ with $\mu_\mu$ is much steeper than that of $Y_{\nu_\mu}$ with $\mu_{\nu_\mu}$. 
Therefore, similar increases of $\mu_\mu$ and $\mu_{\nu_\mu}$ due to chemical equilibrium result in 
an increase of $Y_\mu$ much larger than that of $Y_{\nu_\mu}$.
\label{fig:chempot}}
\end{figure*}

As shown in Figs.~\ref{fig:yevol7.1} and \ref{fig:yevol10},
there is little evolution of $Y_e$, the evolution of $Y_{\nu_e}$ is less than that of $Y_{\nu_\mu}$,
and the evolution of $Y_\mu$ is the largest. This comparison is reflected by changes of the corresponding chemical potentials.
For illustration, we show in the left (right) panel of Fig.~\ref{fig:chemev} the evolution of the chemical 
potentials $\mu_e$, $\mu_{\nu_e}$, $\mu_\mu$, and $\mu_{\nu_\mu}$ for the zone at $r=14.1$~km for $m_s=7.1$ (10)~keV.
It can be seen that the changes $|\Delta \mu_e|$ and $|\Delta\mu_{\nu_e}|$ are much smaller than $|\Delta\mu_{\nu_\mu}|$ and $|\Delta\mu_\mu|$,
which along with $(\mu_e-\mu_{\nu_e})-(\mu_\mu-\mu_{\nu_\mu})=0$ for chemical equilibrium [see Eqs.~(\ref{eq:munuenpe}) and (\ref{eq:munumunpmu})],
gives $\Delta\mu_\mu\sim\Delta\mu_{\nu_\mu}$.
For the conditions in the PNS, the increase of $Y_\mu$ with $\mu_\mu$ is much steeper
than that of $Y_{\nu_\mu}$ with $\mu_{\nu_\mu}$ because of the muon rest mass
(see Fig.~\ref{fig:chempot}). Therefore, the modest increase of $Y_{\nu_\mu}$ mainly due to
the MSW conversion of $\bar\nu_\mu$ into $\bar\nu_s$ gives rise to the substantial increase of $Y_\mu$
through chemical equilibrium.

\begin{figure*}[!t]
\centering
\includegraphics[width=0.4\textwidth]{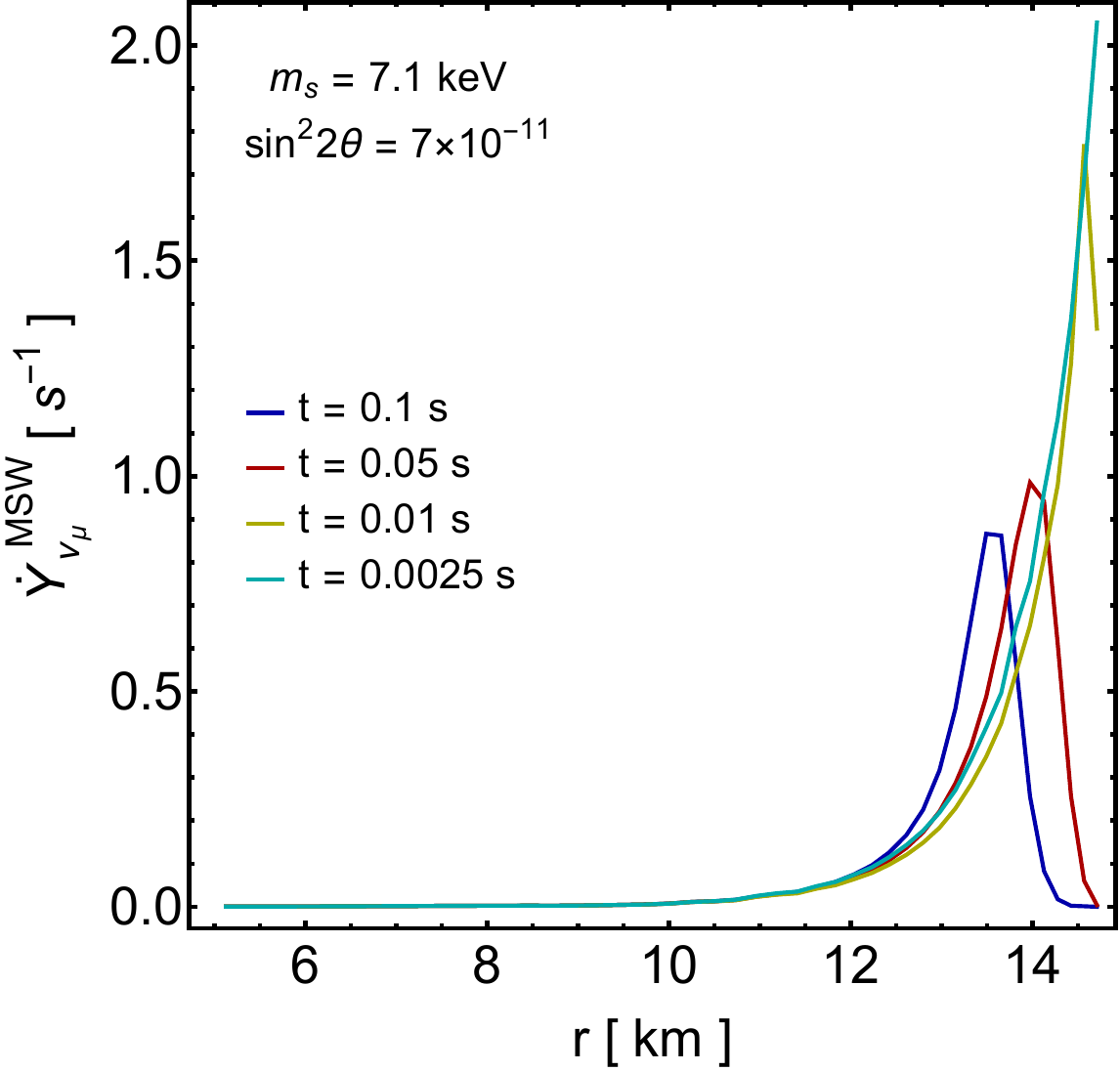}	
\hspace*{0.7 cm}
\includegraphics[width=0.4\textwidth]{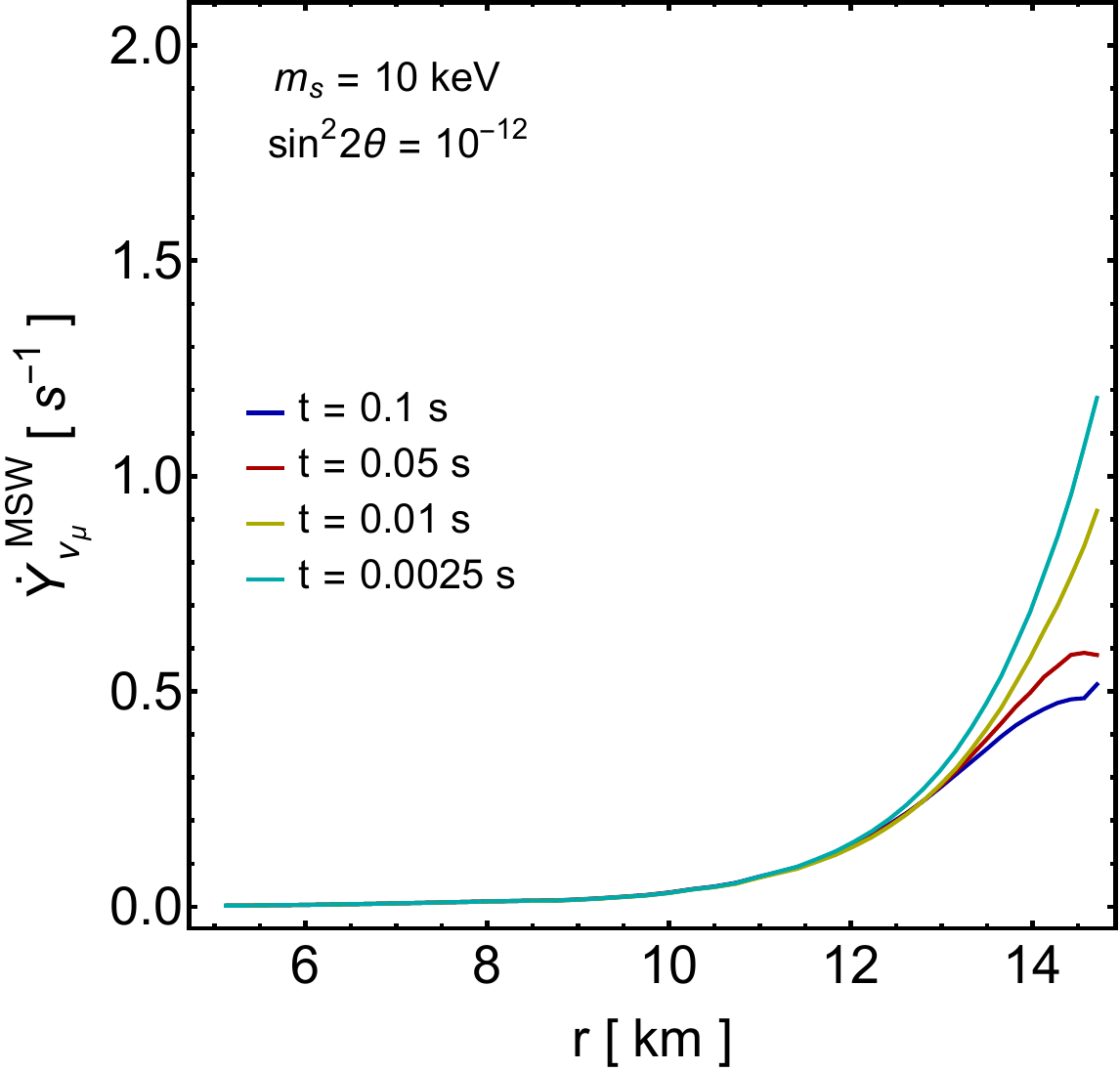}\bigskip\\ 
\includegraphics[width=0.4\textwidth]{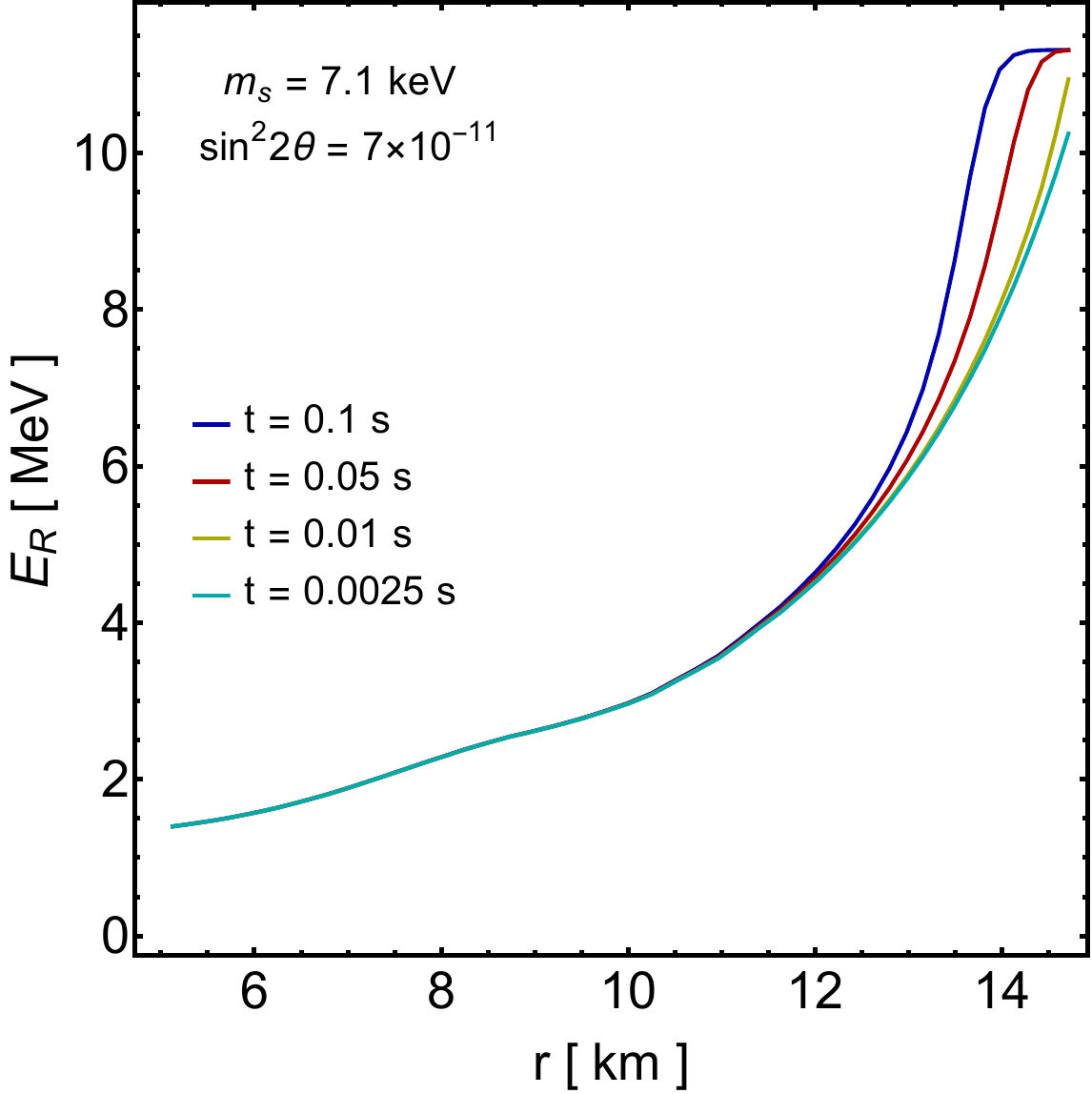}	
\hspace*{0.7 cm}
\includegraphics[width=0.4\textwidth]{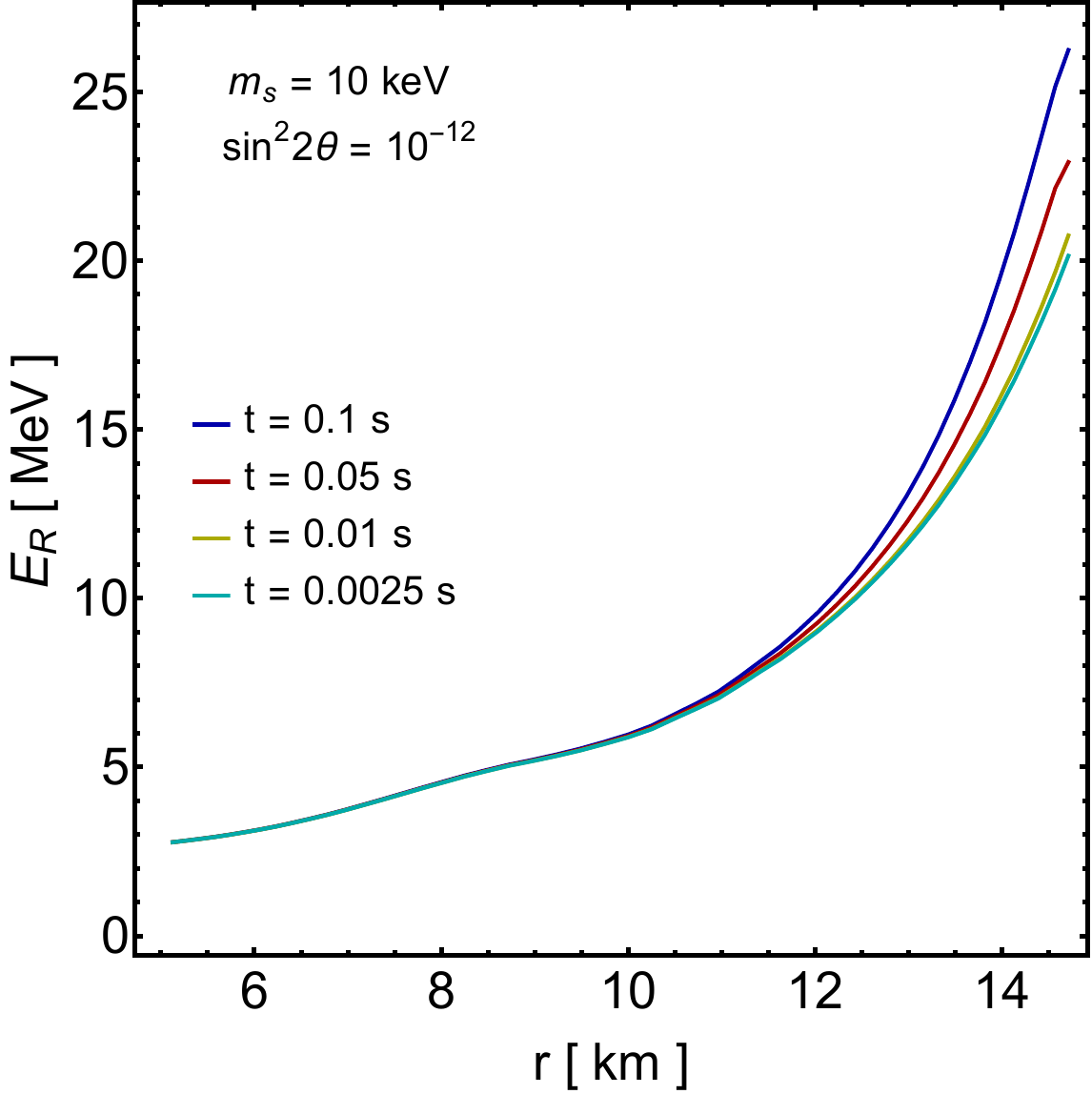}
\caption{Evolution of the rate $\dot Y_{\nu_\mu}^{\rm MSW}$ and the resonant energy $E_R$ 
as functions of radius $r$ and time $t$ for the MSW conversion of $\bar\nu_\mu$ into $\bar\nu_s$
with $(m_s/{\rm keV},\sin^22\theta)=(7.1,7\times10^{-11})$ (left panels) and
$(10,10^{-12})$ (right panels). Diffusion of $Y_{\nu_e}$ and $Y_{\nu_\mu}$ is ignored. \label{fig:ymswer}}
\end{figure*}

To understand better the dominant production of sterile
neutrinos through the MSW conversion of $\bar\nu_\mu$ into $\bar\nu_s$, we show the corresponding rate
$\dot Y_{\nu_\mu}^{\rm MSW}$ for change in $Y_{\nu_\mu}$ as functions of radius $r$ and time $t$ 
in the top panels of Fig.~\ref{fig:ymswer}. Initially $\dot Y_{\nu_\mu}^{\rm MSW}$ increases
with radius as $\bar\nu_\mu$ of higher energy go through MSW resonances at larger radii
(bottom panels of Fig.~\ref{fig:ymswer}). The resulting production and escape of $\bar\nu_s$
change $Y_{\nu_{\mu}}$ directly and alter $Y_\mu$, $Y_e$, $Y_{\nu_e}$, $Y_n$, and $Y_p$ through 
chemical equilibrium. Consequently, the potential $V_{\bar\nu}$ is changed,
which in turn affects the subsequent MSW conversion of $\bar\nu_\mu$ into $\bar\nu_s$. These feedback
effects are clearly shown in Fig.~\ref{fig:ymswer}. At the same radius, the resonant
energy $E_R\approx\delta m^2/(2V_{\bar\nu})$ increases with time, mostly due to the increase of $Y_\mu$ 
and hence decrease of $V_{\bar\nu}\sim \sqrt{2}G_Fn_b[(Y_n/2)-Y_\mu]$. The evolution of 
$\dot Y_{\nu_\mu}^{\rm MSW}$ is more complicated because it depends on both the resonant energy and
the evolving energy distribution of $\bar\nu_\mu$ (due to the change of $\mu_{\nu_\mu}$). 
The general tendency is that $\dot Y_{\nu_\mu}^{\rm MSW}$
decreases with time, i.e., the feedback of $\nu_\mu$-$\nu_s$ and $\bar\nu_\mu$-$\bar\nu_s$ mixing
tends to turn off such mixing (e.g., \cite{2011PhRvD..83i3014R,2019JCAP...12..019S,PhysRevD.108.063025}).
While the cases of $m_s=7.1$ and 10~keV are similar, the relevant resonant energy is higher
(due to a higher $\delta m^2$) and the corresponding MSW conversion is less adiabatic (mostly due to a much 
smaller $\sin^22\theta$) for the latter. Consequently, the evolution of $Y_\mu$ is significantly
different for the two cases (the differences in the evolution of $Y_{\nu_e}$ and $Y_{\nu_\mu}$ are
also noticeable, see Figs.~\ref{fig:yevol7.1} and \ref{fig:yevol10}).

\section{Discussion and conclusions}
\label{sec:discuss}
We have presented the effects of $\nu_{\mu}$-$\nu_{s}$ and $\bar\nu_{\mu}$-$\bar\nu_{s}$ 
mixing with $\delta m^2\sim 10^2$~keV$^2$ in the PNS. In addition to the obvious feedback on the $\nu_\mu$
lepton number, we have included for the first time, the feedback on the composition of the PNS. 
For our adopted mixing parameters, which are consistent
with current constraints, we find that sterile neutrinos are dominantly produced through
the MSW conversion of $\bar\nu_\mu$ into $\bar\nu_s$. This production and the subsequent 
escape of $\bar\nu_s$ increase the $\nu_\mu$ lepton number, which in turn enhances
muonization mainly through $\nu_\mu+n\to p+\mu^-$ and changes the number fractions 
$Y_e$, $Y_{\nu_e}$, $Y_p$, and $Y_n$ through chemical equilibrium.

In our adopted CCSN model \cite{2017PhRvL.119x2702B}, $Y_\mu$ reaches the peak value of 0.0566 at $r=8.97$~km 
at $t_{\rm pb}=1$~s with standard physics. By including $\nu_{\mu}$-$\nu_{s}$ and $\bar\nu_{\mu}$-$\bar\nu_{s}$ mixing 
with $\delta m^2\approx m_s^2\sim 10^2$~keV$^2$, we have shown that enhanced muonization occurs in that model at larger 
radii ($r\sim 12$--14.7~km) beyond the original peak of $Y_\mu$ (Figs.~\ref{fig:yevol7.1} and \ref{fig:yevol10}). This enhancement is
driven by the MSW resonances of $\bar\nu_\mu$ with $E_R\sim 4$--10 (8--20)~MeV for $m_s=7.1$ (10)~keV (Fig.~\ref{fig:ymswer}).
We have ignored diffusion of $Y_{\nu_e}$ and $Y_{\nu_\mu}$ in our calculations. 
Based on our adopted CCSN model with standard physics but with detailed treatment of such diffusion,
the PNS is contracting between $t_{\rm pb}=1$ and 1.25~s and the resulting change in the density profile and more importantly, 
that in the temperature profile are most likely the dominant factors driving the evolution of composition profiles (see Fig.~\ref{fig:cond}). 
However, changes of $Y_{\nu_e}$, $Y_{\nu_\mu}$, and especially $Y_\mu$ found in our calculations 
(assuming fixed temperature and density profiles over a duration of 0.1~s)
are qualitatively different from those in the CCSN model with standard physics (see Figs.~\ref{fig:yevol7.1} and \ref{fig:yevol10}). 
Further, in the region of $r\sim 12$--14.7~km, the increase of $Y_\mu$ obtained from our calculations quantitatively exceeds 
the corresponding change in the standard CCSN model.
Therefore, we regard enhanced muonization as a qualitatively robust result from $\nu_{\mu}$-$\nu_{s}$ and $\bar\nu_{\mu}$-$\bar\nu_{s}$ 
mixing with $\delta m^2\sim 10^2$~keV$^2$ in the PNS, and expect that it still occurs when such mixing is incorporated in CCSN
simulations. Such self-consistent simulations should be carried out to assess quantitatively the effects discussed here.

We have chosen to study the epoch of $t_{\rm pb}\approx 1$~s because the corresponding temperature and density profiles 
change so slowly that we can ignore such changes when studying the effects of $\nu_{\mu}$-$\nu_{s}$ and 
$\bar\nu_{\mu}$-$\bar\nu_{s}$ mixing for a duration of 0.1~s. On the other hand, the PNS conditions do not
differ greatly for $t_{\rm pb}\sim 0.4$--2~s and the conditions at $t_{\rm pb}\sim 0.1$~s are also not qualitatively different, 
which suggests that our results can be extended to earlier and later times.
With the help of muonization, neutrino-driven explosion occurs at $t_{\rm pb}\sim 0.24$~s
in our adopted CCSN model \cite{2017PhRvL.119x2702B}. We expect that
when $\nu_{\mu}$-$\nu_{s}$ and $\bar\nu_{\mu}$-$\bar\nu_{s}$ mixing with $\delta m^2\sim 10^2$~keV$^2$ is 
included in CCSN simulations, enhanced muonization similar to that found here would occur at the
explosion epoch, potentially making the explosion easier. However, due to the intrinsically dynamic nature of 
the explosion, a definitive conclusion can only be reached by implementing such mixing in CCSN simulations.

Subsequent to the explosion, cooling of the PNS by active neutrino emission lasts up to $\sim 10$~s.
Active neutrinos emitted from the PNS drive winds (e.g., \cite{1996ApJ...471..331Q}) that
may be an important source for heavy element nucleosynthesis (e.g.,\,\cite{1997ApJ...482..951H}). 
Further, this nucleosynthesis may be affected by collective
oscillations of active neutrinos (see e.g., \cite{FISCHER2024104107} for a recent review) that are
sensitive to the net lepton number carried by $\nu_\mu$ and $\bar\nu_\mu$ fluxes \cite{PhysRevLett.125.251801}.
With enhanced muonization due to $\nu_{\mu}$-$\nu_{s}$ and $\bar\nu_{\mu}$-$\bar\nu_{s}$ mixing in the PNS,
we expect that a significant net $\nu_\mu$ lepton number is emitted from the PNS, but note that the quantitative
characteristics of active neutrino emission can only be obtained by including such mixing in CCSN simulations. 
We will explore the implications of enhanced muonization for the CCSN dynamics and collective neutrino oscillations 
in future works.
\section*{ACKNOWLEDGMENT}
We sincerely thank Daniel Kresse and Thomas Janka for providing the CCSN model used in this work. 
This work was supported in part by the National Science Foundation (Grant No. PHY-2020275), 
the Heising-Simons Foundation (Grant 2017-228), and the U.S. Department of Energy (Grant No. DE-FG02-87ER40328).
\appendix*
\section{Charged-current weak reaction rates}
Under the so-called elastic approximation (the nucleons carry the same momentum), the rate for $\nu_e+n\to p+e^-$ is given by
\begin{align}
\Gamma_{\nu_e n}&= \frac{1}{n_bY_n}\int_{0}^{\infty}\frac{p^2dp}{\pi^2}f_n(E_n)[1-f_p(E_p)]\nonumber\\
&\times\int_0^\infty\frac{E_{\nu_e}^2dE_{\nu_e}}{2\pi^2}f_{\nu_e}(E_{\nu_e})[1-f_{e^{-}}(E_{e^-})]\sigma_{\nu_en}(E_{e^-}), 
\end{align}
where
\begin{align}
    \sigma_{\nu_en}(E_{e^-})=\frac{G^2_F}{\pi}\cos^2 \theta_c(f^2+3g^2)E_{e^-} p_{e^-}
    \label{eq:xsnuen}
\end{align}
is the cross section, $\theta_c$ denotes the Cabibbo angle with $\cos^2 \theta_c\approx0.95$,
$f=1$ and $g\approx1.27$ are weak coupling constants,
\begin{align}
    f_\alpha(E_\alpha)=\frac{1}{\exp[(E_\alpha-\mu_\alpha)/T]+1}\,
\end{align}
is the occupation number of species $\alpha$ with $\alpha=n$, $p$, $\nu_e$, and $e^-$, and 
the chemical potential $\mu_\alpha$ includes the rest mass $m_\alpha$ of the species.
In the above equations,
\begin{align}
    E_{n,p}&=\sqrt{p^2+(m_{n,p}^*)^2}-m_{n,p}^*+m_{n,p}+U_{n,p},\\
    E_{e^-}&=\sqrt{p_{e^-}^2+m_e^2}\\
    &=E_{\nu_e}+m_n-m_p+U_n-U_p,
\end{align}
where for example, $m_n^*$ and $U_n$ are the effective mass and potential, respectively, for neutrons in the PNS
(e.g.,\,\cite{FISCHER2024104107}).

The rate $\Gamma_{\nu_\mu n}$ for $\nu_{\mu}+n\to p+\mu^-$ can be obtained by replacing quantities
for $\nu_e$ ($e^-$) with those for $\nu_\mu$ ($\mu^-$) in the equations for $\Gamma_{\nu_e n}$.

The rate for $\bar\nu_e+p\to n+e^+$ is given by
\begin{align}
\Gamma_{\bar{\nu}_e p}&= \frac{1}{n_bY_p}\int_{0}^{\infty}\frac{p^2dp}{\pi^2}f_p(E_p)[1-f_n(E_n)]\nonumber\\
&\times\int_{E_{\bar\nu_e}^{\rm th}}^\infty\frac{E_{\bar{\nu}_e}^2dE_{\bar{\nu}_e}}{2\pi^2}f_{\bar{\nu}_e}(E_{\bar{\nu}_e})[1-f_{e^{+}}(E_{e^+})]\sigma_{\bar{\nu}_ep}(E_{e^+}),
\end{align}
where $\sigma_{\bar{\nu}_ep}(E_{e^+})$ can be obtained by replacing $E_{e^-}$ ($p_{e^-}$) with $E_{e^+}$ ($p_{e^+}$) in Eq.~(\ref{eq:xsnuen})
for $\sigma_{\nu_en}(E_{e^-})$,
\begin{align}
    E_{e^+}=E_{\bar{\nu}_e}+m_p-m_n+U_p-U_n,
\end{align}
and $E_{\bar\nu_e}^{\rm th}$ corresponds to $E_{e^+}=m_e$. Note that in the occupation number $f_{\bar\alpha}(E_{\bar\alpha})$ for an antiparticle $\bar\alpha$,
the chemical potential is $\mu_{\bar\alpha}=-\mu_{\alpha}$.

The rate $\Gamma_{\bar{\nu}_{\mu} p}$ for $\bar\nu_{\mu}+p\to n+\mu^+$ can be obtained by replacing quantities
for $\bar\nu_e$ ($e^+$) with those for $\bar\nu_\mu$ ($\mu^+$) in the equations for $\Gamma_{\bar\nu_e p}$.

By detailed balance, the rates for $e^++n \to p + \bar{\nu}_e $, $\mu^++n\to p+\bar\nu_{\mu}$,
$e^-+p \to n + \nu_e $, and $\mu^-+p \to n + \nu_{\mu}$ are given by
\begin{align}
\Gamma_{e^+n}&=\frac{Y_p}{Y_n}\Gamma_{\bar{\nu}_e p}\exp\left(\frac{\mu_{\nu_e}+\mu_n-\mu_p-\mu_e}{T}\right),\\
\Gamma_{\mu^+ n}&=\frac{Y_p}{Y_n}\Gamma_{\bar{\nu}_{\mu} p}\exp\left(\frac{\mu_{\nu_\mu}+\mu_n-\mu_p-\mu_\mu}{T}\right),\\
\Gamma_{e^-p}&=\frac{Y_n}{Y_p}\Gamma_{\nu_e n}\exp\left(\frac{\mu_e+\mu_p-\mu_n-\mu_{\nu_e}}{T}\right),\\
\Gamma_{\mu^-p}&=\frac{Y_n}{Y_p}\Gamma_{\nu_\mu n}\exp\left(\frac{\mu_\mu+\mu_p-\mu_n-\mu_{\nu_\mu}}{T}\right).
\end{align}
The exponential factors in the above equations reduce to unity when the relevant particles are in chemical equilibrium.
We have checked that the charged-current weak reaction rates are sufficiently fast, and therefore, chemical equilibrium
is achieved to very good approximation.
\bibliographystyle{JHEP}
\bibliography{ref}
\end{document}